\documentclass[prd,nofootinbib,preprint,floatfix]{revtex4}
\usepackage{amsmath,amssymb}
\usepackage{epsfig}
\usepackage{axodraw}
\usepackage[dvips]{color}

\newcommand{\sla}[1]{/\!\!\!#1}
\def\beq{\begin{equation}}
\def\eeq{\end{equation}}
\def\beqn{\begin{eqnarray}}
\def\eeqn{\end{eqnarray}}
\def\lsim{\ ^<\llap{$_\sim$}\ }

\def\m32{$m_{3/2}$}
\def\m0{$m_0$}

\def\sec2w{sec^2\theta_W}
\def\12{$\frac{1}{2}$}
\begin{document}

\title{LHC signals for neutrino mass model in \\
bilinear R--parity violating mAMSB}

\vskip 1.5 true cm
\author{F.\ de Campos}
\email{camposc@feg.unesp.br}
\affiliation{Departamento de F\'{\i}sica e Qu\'{\i}mica,
             Universidade Estadual Paulista, Guaratinguet\'a, SP,  Brazil }

\author{M.\ A.\ D\'iaz}
\email{mad@susy.fis.puc.cl}
\affiliation{Departamento de F\'{\i}sica,
 Universidad Cat\'olica de Chile,
 Av. V. Mackenna 4860, Santiago, Chile.}

\author{O.\ J.\ P.\ \'Eboli}
\email{eboli@fma.if.usp.br}
\affiliation{Instituto de F\'{\i}sica,
 Universidade de S\~ao Paulo, S\~ao Paulo, SP, Brazil.}

\author{M.\ B.\ Magro}
\email{magro@fma.if.usp.br}
\affiliation{Faculdade de Engenharia, CUFSA
  Santo Andr\'e, Brazil.}

\author{W.\ Porod}
\email{porod@physik.uni-wuerzburg.de}
\affiliation{Institut f\"ur Theoretische Physik and Astrophysik, Universit\"at
W\"urzburg,  D-97074 W\"urzburg, Germany}

\author{S. Skadhauge}
% \email{solveig@fma.if.usp.br}
% \affiliation{Instituto de F\'{\i}sica, Universidade de S\~{a}o Paulo,
% S\~{a}o Paulo, SP, Brazil}
\email{solveig@nordita.org}
\affiliation{Nordita, AlbaNova University Center, Roslagstullbacken 23,
SE-10691 Stockholm, Sweden}

%%%%%%%%%%%%%%%%%%%%%%%%%%%%%%%%%%%%%%%%%%%%%%%%%%%%%%%%%%%%
\begin{abstract}
%\vskip 0.5 true cm

  We investigate a neutrino mass model in which the neutrino data is
  accounted for by bilinear R--parity violating supersymmetry with
  anomaly mediated supersymmetry breaking. We focus on the CERN Large
  Hadron Collider (LHC) phenomenology, studying the reach of generic
  supersymmetry search channels with leptons, missing energy and jets.
  A special feature of this model is the existence of long lived
  neutralinos and charginos which decay inside the detector leading to
  detached vertices.
  We demonstrate that the largest reach is obtained in the displaced
  vertices channel and that practically all of the reasonable
  parameter space will be covered with an integrated luminosity of 10
  fb$^{-1}$. We also compare the displaced vertex reaches of the LHC 
  and Tevatron.

\end{abstract}

\maketitle
%%%%%%%%%%%%%%%%%%%%Section%%%%%%%%%%%%%%%%%%%%%%%%%%%%%%%%%
\section{Introduction}\label{intro}

The neutrino sector is one of the most exciting sectors of particle
physics today. Our knowledge of the neutrino parameters has increased
tremendously during the past decade. Both the atmospheric and the
solar mass squared differences and mixing angles are known to a good
precision~\cite{atm,solar,kamland,k2k,minos}. However, only an upper
bound exists on the so-called CHOOZ angle, $\sin^2\theta_{13}<0.04$
\cite{chooz}, and the absolute mass scale for the neutrino masses,
$\sum m_{\nu}\lsim 0.6$ eV~\cite{wmap}.  Furthermore, there is an
ambiguity in the sign of the atmospheric mass square difference
leading to the possibility of two different hierarchies, normal or
inverse, for the neutrino masses.

As the neutrino experiments enter a precision phase the need to
explore different neutrino mass models increases. The most popular
model for neutrino masses, the seesaw mechanism~\cite{seesaw},
beautifully explains the smallness of the neutrino masses as compared
to other fermion masses. Nevertheless, it is difficult to test this
mechanism due to its very high energy scale.  The right-handed
neutrinos introduced in the seesaw model have masses of order
$10^{12}$ GeV and are too heavy to be produced in
colliders\footnote{However, one might see there traces in the 
properties of the left sneutrinos~\cite{FPZ}}.  On the other hand,
models where the origin of neutrino masses is related to TeV scale
physics have been proposed  \cite{Zee:1980ai,Babu:1988ki} 
and these might be tested at the future or even present 
colliders  \cite{AristizabalSierra:2006ri,AristizabalSierra:2006gb}.

In this paper we will consider a TeV--scale mechanism for generating
neutrino masses, namely through R--parity violating (RPV)
supersymmetry (SUSY)~\cite{rpvreview}. In this scenario the neutrino
and neutralinos mix giving rise to the neutrino masses~\cite{susyneu}.
We will be constraining ourself to bilinear RPV
(BRpV)~\cite{bili,bili2}, thus breaking lepton number but not baryon
number. This model can be viewed as the effective theory of a
spontaneously broken R--parity symmetry~\cite{spon}. The BRpV neutrino
mass model has only a few free parameters and therefore is very
predictive.  Furthermore, in contrast to trilinear RPV neutrino mass
models, the constraints from LEP on the R--Parity violating couplings
are automatically satisfied as the couplings are small.

We will study the specific case of anomaly mediated supersymmetry
breaking (AMSB)~\cite{Giudice,Randall}.  In this scenario the
contribution to the soft-supersymmetry breaking terms from the
super-conformal anomaly, which is always present, is assumed to be
dominant.  AMSB naturally solves the flavor problem of the minimal
supersymmetric standard model (MSSM), since the masses of the two
first generations of scalars are automatically equal and the flavor
off-diagonal terms are given in terms of the quark Yukawa couplings.
However, anomaly mediation in its pure form is not a viable theory
since without any other soft symmetry breaking terms, tachyonic
sleptons are present in the particle spectrum.  Therefore, one
normally adds an universal scalar mass term.  This minimal anomaly
mediated scenario (mAMSB) is the one we will pursue in an extended
form where we also add bilinear R--parity violation.

It has been shown that mAMSB with BRpV can account for the present
neutrino data~\cite{DeCampos:2001wq,deCampos:2004iy}.
Like any SUSY BRpV neutrino mass model it predicts the
normal hierarchy, since the neutrino masses becomes strongly
hierarchical. The hierarchy is induced because only one neutrino mass
is generated at tree-level and the other masses are generated at
loop level. In general the atmospheric mass squared difference and
mixing angle are related to tree level physics, whereas the solar
mass squared difference and mixing angle are established by
radiative corrections. The introduction of RPV will render
the lightest supersymmetric particle (LSP), which in mAMSB is
normally the lightest neutralino, unstable. Clearly, this fact
will be very important for the collider phenomenology. Indeed,
the standard signal with much missing energy expected from
R--parity conserving (RPC) supersymmetry will be depleted.

The RPV couplings giving rise to neutrino masses are also responsible
for the decay properties of the neutralino.  Therefore, an important
smoking gun signal for these models is the strong connection between
neutralino physics and neutrino mixing parameters.  Some ratios of the
branching ratios of the neutralino are related to neutrino mixing
angles~\cite{rpvcol,porod1}.  In particular, approximately the same
number of muon as taus are expected along with a W-boson because their
ratio is given by tangent of the nearly maximal atmospheric mixing
angle. By measuring the decay properties of the neutralino, which is
likely to be done by the LHC, a severe test of this model is possible.

Due to the smallness of the RPV couplings the neutralino will have a
long lifetime, but short enough for it to mainly decay within the
detectors at the LHC.  A distinct feature of the mAMSB scenario is the
near degeneracy of the lightest neutralino and the lightest chargino,
causing even the chargino to dominantly decay through RPV couplings.
Consequently, also the chargino will have a lifetime in the range
interesting for colliders such that it will travel a macroscopic
distance but decay before leaving the inner detector.  A very
interesting and unique signal will therefore be the observation of
displaced vertices from the neutralino and the chargino. Displaced
vertices may also be produced, although only arising from neutralino,
in the minimal supergravity (mSUGRA) scenario and has been shown to give
an excellent reach of the model parameters~\cite{deCampos:2005ri}. As
the mAMSB has two different long-lived sparticles, there is a 
richer set of possible final states.

The prospects for collider discovery of RPV, responsible for the
neutrino masses and mixings, in mSUGRA have been thoroughly
studied~\cite{rpvcol,porod1,barger, Magro:2003zb,deCampos:2005ri}.
Also, the discovery prospects for mAMSB with R--parity conservation at
colliders have been
analyzed~\cite{Baer:2000bs,Barr:2002ex,Roy:2004xm}. Nevertheless, the
collider signals within R--parity violating mAMSB scenario have been
analyzed so far only considering trilinear R-parity
violation~\cite{Allanach:2000gu}.
In this paper we will study the BRpV-mAMSB scenario, requiring that
the neutrino masses and mixings are generated by the RPV couplings.
We will analyze various generic SUSY search channels for the LHC.  In
addition we will also determine the reach in the displaced vertices
channel.

We organize the paper as follows.  In section \ref{model} we will
review the model and the main low energy constraints. In section
\ref{signals} we outline our choice of final state channels and
describe our simulation of the signals and backgrounds. In section
\ref{results} we present our results, as well as, our conclusions.

%%%%%%%%%%%%%%%%%%%%Section%%%%%%%%%%%%%%%%%%%%%%%%%%%%%%%%%
\section{ANOMALY MEDIATED SUPERSYMMETRY WITH BRPV}
\label{model}

In this section we will give a short review of the model we analyze.
For further details we refer to \cite{deCampos:2004iy}.  The R--parity
conserving soft terms will be assigned according to the minimal
anomaly mediated scenario. In addition we will allow for bilinear
R--parity violation which contributes in total with six more
parameters.  The R--parity violating couplings are restricted to
values which are consistent with the neutrino data, which severely
constraints the available parameter space.

%%%%%%%%%%%%%%%%%%%%%%%%%%%Figure%%%%%%%%%%%%%%%%%%%%%%%%%%%%%%%%%
\begin{figure}[t]
\centering
\includegraphics[width=10cm,height=8.0cm]{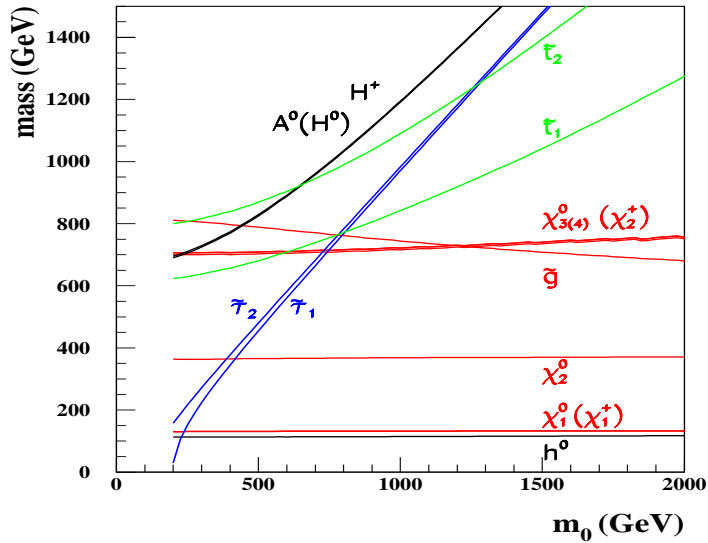}
\caption{Various sparticle masses for $\tan\beta=10$, $\mu>0$, and
$m_{3/2}=40$ TeV as a function of  $m_{0}$.}
\label{fig:spec}
\end{figure}
%%%%%%%%%%%%%%%%%%%%%%%%%%%Figure%%%%%%%%%%%%%%%%%%%%%%%%%%%%%%%%%

Let us start by describing the mAMSB model.
This model can be parameterized by three parameters and a sign,
\begin{equation}\label{paramsb}
 m_{3/2} , m_0 , \tan \beta, {\rm sign}(\mu)
\end{equation}
all defined at the high energy scale $M_{\rm GUT}$.  The gravitino
mass, $m_{3/2}$ , is much larger than the other masses, as this is the
only one generated at tree level.  All the soft breaking terms are
proportional to the gravitino mass in pure AMSB. However, as mentioned
above, in order to avoid tachyonic slepton masses an extra universal
scalar mass $m_0^2$ is added to all sfermions and Higgs masses in
mAMSB. The ratio between the vacuum expectation values of $H_u$ and
$H_d$ is as usual denoted by $\tan \beta$.  Finally, the sign of the
Higgs/Higgsino mixing parameter, $\mu$, is free.  Our conventions are
such that $\mu$ enters the superpotential as $-\mu \hat{H}_d \hat{H}_u
$.  For the explicit relations between the input parameters in
(\ref{paramsb}) and the soft breaking terms in mAMSB see {\it
  e.g.}~\cite{Gherghetta:1999sw}.

A typical spectrum for the anomaly mediated scenario is shown in
Figure~\ref{fig:spec}.  Throughout most of the viable parameter space
the lightest neutralino is the LSP, with only a small area having the
stau or the tau--sneutrino as the LSP. Whether the stau is heavier or
lighter than the tau--sneutrino depends on the value of $\tan\beta$.
For sufficient high values of $m_0$ the lightest neutralino will be
the LSP.  The gaugino masses are proportional to their beta functions,
resulting in the unique relationship for AMSB: \ $M_1 : M_2 : M_3 \sim
3 : 1 : 7$.  Here $M_i$ are the gaugino masses; $M_1$ is the bino
mass, $M_2$ the wino mass and $M_3$ the gluino mass.  This explains
the first of two distinct characteristics of the mAMSB spectrum; the
near degeneracy of the lightest neutralino (also the LSP is most
cases) and the lightest chargino. As the wino mass is much lighter
than the others it will be almost equal to the masses of the lightest
states, with the neutral state being slightly lighter. The wino--like
nature of the lightest neutralino is important as its interactions are
stronger and it will be more easily produced at colliders.  The second
characteristic, the near degeneracy of the sleptons, is a less robust
feature, as it is a consequence of the assumption of a universal extra
contribution to the scalar masses and thus a feature of the minimal
AMSB.

The invariance under R--parity, defined by $R_p = (-1)^{3(B-L)+2S}$,
is normally assumed in supersymmetric models.  This is mainly
motivated by two requirements: to obtain a stable proton and to get
the LSP as a dark matter candidate. However, the first requirement can
also be obtained by different symmetries forbidding only baryon number
violating terms; see, {\em e.g.},~\cite{Dreiner:2005rd}. Introducing
lepton number violating terms has the benefit that neutrino masses are
generated in an intrinsically supersymmetric way.
We will introduce
bilinear R--parity violation in order to produce the observed
neutrino masses and mixings. Thus, we add the following term
\begin{equation}
 W_{BRpV} = \epsilon_i \hat{L}_i  \hat{H}_u \,, \;\;\;\; i = 1, \ldots , 3
\label{eq:bilinear}
\end{equation}
to the MSSM superpotential. In order to acquire agreement with the
neutrino data, the R�-parity violating couplings must satisfies
$\epsilon_i \ll  \mu$.
For consistency also the soft breaking terms,
\begin{equation}
B_i \epsilon_i L_i H_u \;,\;\;\;\; i = 1, \ldots, 3
\label{eq:softbilinear}
\end{equation}
are added to the MSSM. Thus, six parameters related to break down of
R--parity invariance are introduced. In general this will give rise to
sneutrino vacuum expectation values (vevs), which in turn leads to
mixing between neutralinos and neutrinos. The four heaviest states
from the diagonalization of the $7 \times 7$ neutralino--neutrino mass
matrix will be almost pure neutralino states and we denote these by
$\chi^0_k \;, \; k = 1, \ldots , 4$. Moreover, we arrange them in
order of magnitude of the masses, thus, $\chi^0_1$ is the lightest
neutralino. The chargino--charged lepton mass matrix is treated in an
analog way and the lightest chargino is denoted by $\chi^{+}_1$.

Clearly the introduction of RPV has important consequences for the
collider phenomenology as it will render the LSP unstable.  As
mentioned above, the lightest supersymmetric particle is normally the
lightest neutralino in mAMSB as is the case in mSUGRA. As we allow for
R--parity violation, also the areas with stau or sneutrino LSP are
viable, these particles will decay.  Nevertheless, since the stau or
sneutrino LSP parameter space regions are very small, we will not
discuss these areas further although they are properly included in our
analysis.  In the following discussion we will assume that $\chi_1^0$
is indeed the LSP.

As mentioned above, the R�-parity conserving parameters are calculated
as in the mAMSB scenario. Thus, these are fixed at the scale $M_{\rm
  GUT}$ and renormalization group equation (RGE) running is used in
order to extract the low energy parameters. We will use the program
SPheno~\cite{spheno}, suitably expanded to the case of RPV, for
calculating the mass spectrum and decay widths.  For the RPC
parameters we use 2-loop RGE's and we include all 1-loop threshold
corrections. The unification scale is defined as the scale where the
$U(1)$ and the $SU(2)$ coupling constants meet.  The RPV couplings are
only dealt with at the low energy scale.  In the case of the bilinear
parameters in the superpotential this can be done consistently without
any additional assumption as also the modulus of $\mu$ is calculated
at the electroweak scale and the bilinear parameters form a closed
system within the RGE evolution~\cite{twoloop}.  The corresponding
soft SUSY breaking parameters also form a closed system~\cite{twoloop} 
but one has to assume that there are additional contributions at 
the high scale similar to the case of the scalar mass
parameters squared to get a consistent picture.

The parameters of eqs.~(\ref{eq:bilinear}) and (\ref{eq:softbilinear})
are determined with the help of neutrino physics. First we trade the
$B_i$ by the sneutrino vevs $v_i$ using the tadpole equations. The
neutrino masses and mixings are best parameterized using the quantities
$\Lambda_i = \mu v_i + \epsilon_i v_d$ and $\tilde \epsilon_i = V_{ij}
\epsilon_j$ where $V_{ij}$ is the mixing matrix of the tree level
neutrino mass matrix \cite{bili2}. In case the tree-level contribution
dominates in the effective neutrino mass matrix, the modulus of $\vec
\Lambda$ is fixed by requiring the correct atmospheric neutrino mass
difference squared, the atmospheric neutrino mixing angle, and the
Chooz angle within the allowed experimental range. In addition, the
ratios $|\vec \epsilon|^2 / |\vec \Lambda|$ and $\tilde \epsilon_1 /
\tilde \epsilon_2$ are fixed by requiring the correct solar mass
difference and solar mixing angle within the experimental range,
respectively.

The very particular near degeneracy of $\chi_1^0$ and $\chi^+_1$ is
preserved in our model, as the R�-parity violating couplings have
little impact on the sparticle masses. It is important to calculate
the chargino and neutralino masses very precisely, as the mass
splitting between these particles, $\Delta m_{\chi} =
m_{\chi^0_1}-m_{\chi^+_1}$, is very small and the exact value can have
important consequences for the chargino lifetime. In our numerical
evaluation, the neutrino�-neutralino and chargino--charged--lepton
mass matrices are evaluated to 1-loop order.

%%%%%%%%%%%%%%%%%%%%%%%%%%%%%%%%%%%%%%%%%%%%%%%%%%%%%%%%%%%%
\subsection{Constraints on the model}

Here we will discuss the existing constraints from low energy
observables. Besides the requirement of agreement with neutrino data,
there are also important constraints from the LEP data, the rare
process $b \rightarrow s \gamma$ and the anomalous magnetic moment of
the muon.

We use the following neutrino constraints, which are the present
ones at 90\% confidence level:
\begin{eqnarray}
+7.3 \times 10^{-5} {\rm eV^2} <
& \Delta m^2_{\rm sol} &
< +9.0 \times 10^{-5} {\rm eV^2} \nonumber \\
0.25 < & \sin^2 \theta_{\rm sol}& < 0.37 \\
1.5 \times 10^{-3} {\rm eV^2} <
& \vert \Delta m^2_{\rm atm} \vert &
< 3.4 \times 10^{-3} {\rm eV^2} \nonumber \\
0.36 < & \sin^2 \theta_{\rm atm} & \leq 0.64
\label{eq:param}
\end{eqnarray}
It is not always possible to succeed in generating the observed
neutrino masses and mixings, and such points will be excluded
from our analysis.

The LEP collider at CERN has already put lower
bounds on some of the supersymmetric sparticle masses.
We have implemented the following constraints from LEP \cite{pgd}:
\begin{eqnarray}
  m_{\tilde{t}} > 95~ {\rm GeV} \;, \;\;\;
  m_{\tilde{b}} > 85~ {\rm GeV} \;, \;\;\;
  m_{\tilde{\tau}} > 79~ {\rm GeV} \;, \;\;\;  \\ \nonumber
  m_{\tilde{\chi}^+} > 95~ {\rm GeV} \;, \;\;\;
  m_{\tilde{\chi}^0} > 42~ {\rm GeV} \;, \;\;\;
  m_{h^0} > 95~ {\rm GeV} \;.
\label{lep}
\end{eqnarray}
We have checked that also the Tevatron bounds on squark and 
gluino masses are satisfied \cite{Abazov:2006bj}.
The limit on the Higgs mass is somewhat optimistic as the bound in the
MSSM is the same as for the SM Higgs throughout most the available
parameter space~\footnote{Only for small pseudo scalar mass can a
  lighter Higgs boson be permitted~\cite{higgsmass}}.  For this reason
we exhibits the contour for $m_{h^0} = 114$ GeV in our plots of the
collider reach.  The lower bound on the chargino mass translates
almost directly to a lower bound on $m_{3/2}$. For $\tan\beta=10$ we
must require $m_{3/2}>30$ TeV in order to satisfy the chargino mass
bound.

%%%%%%%%%%%%%%%%%%%%%%%%%%%Figure%%%%%%%%%%%%%%%%%%%%%%%%%%%%%%%%%
\begin{figure}[thpb]
%\centering
\includegraphics[width=7.5cm,height=8.0cm]{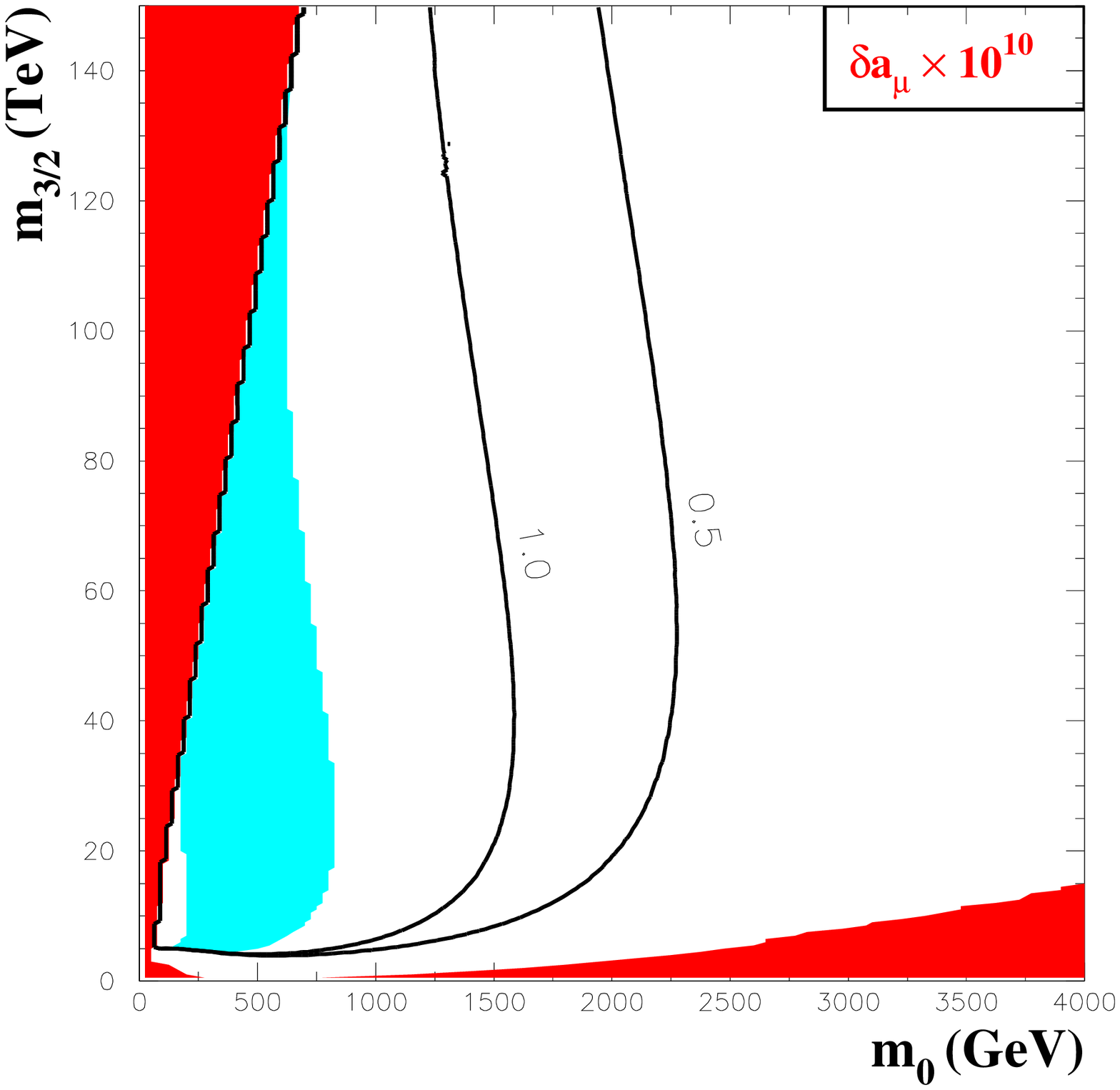}
\includegraphics[width=7.5cm,height=8.0cm]{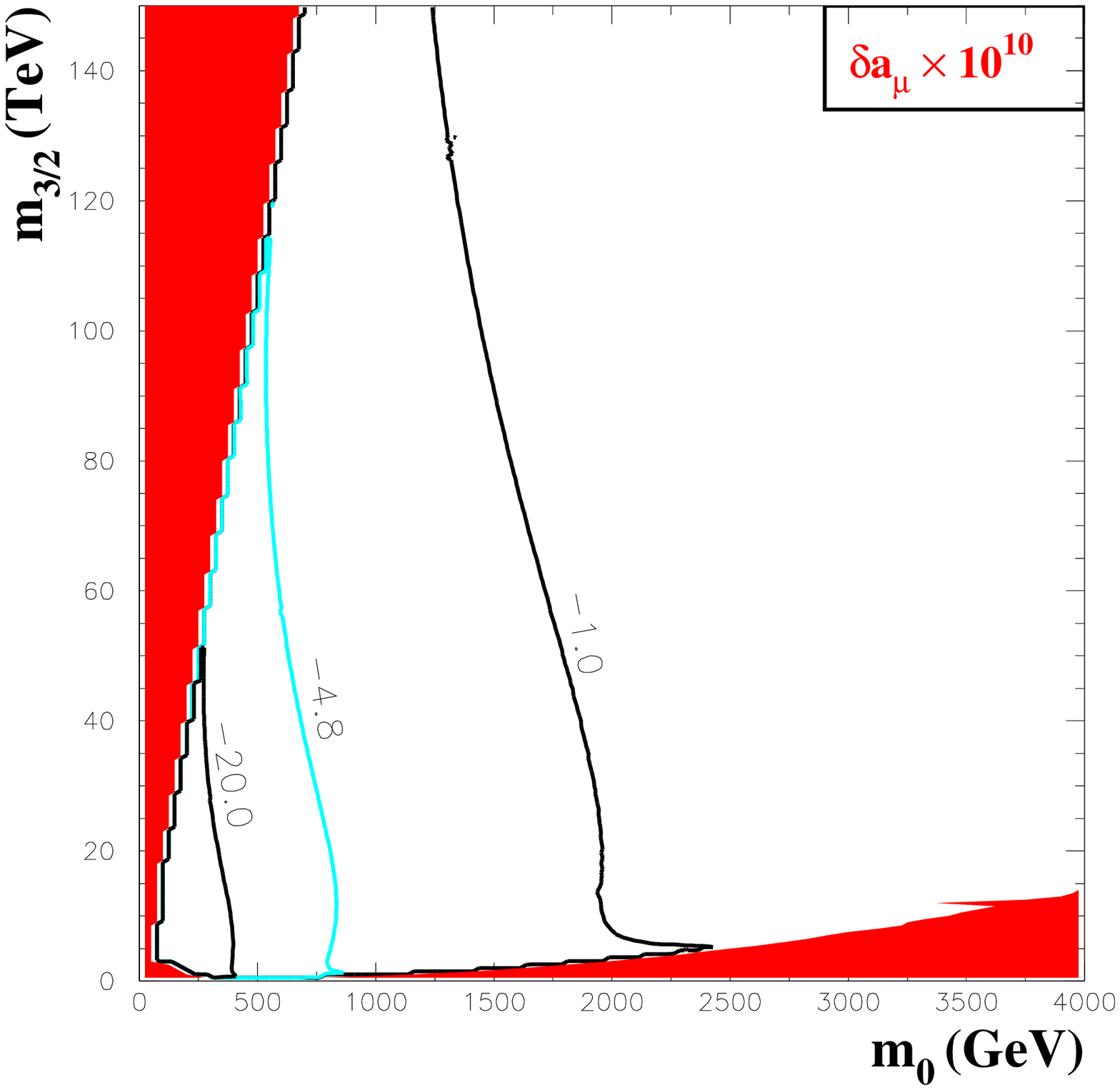}
\caption{The supersymmetric contribution to the anomalous
magnetic moment of the muon for
$\tan\beta=10$ and $\mu>0$ (left), $\mu<0$ (right panel).
The dark (red) area is theoretically excluded, due to either
tachyonic particles or the fact that the electroweak symmetry
is not broken. The light (cyan) area in the left panel is the
allowed 3$\sigma$ range. We also show the contour for
$\delta a_\mu=-4.8 \times 10^{-10}$ in the right panel, which
corresponds to the 4$\sigma$ lower bound.}
\label{fig:g2}
\end{figure}
%%%%%%%%%%%%%%%%%%%%%%%%%%%Figure%%%%%%%%%%%%%%%%%%%%%%%%%%%%%%%%%

%%%%%%%%%%%%%%%%%%%%%%%%%%%Figure%%%%%%%%%%%%%%%%%%%%%%%%%%%%%%%%%
\begin{figure}[thpb]
%\centering
\includegraphics[width=7.5cm,height=8.0cm]{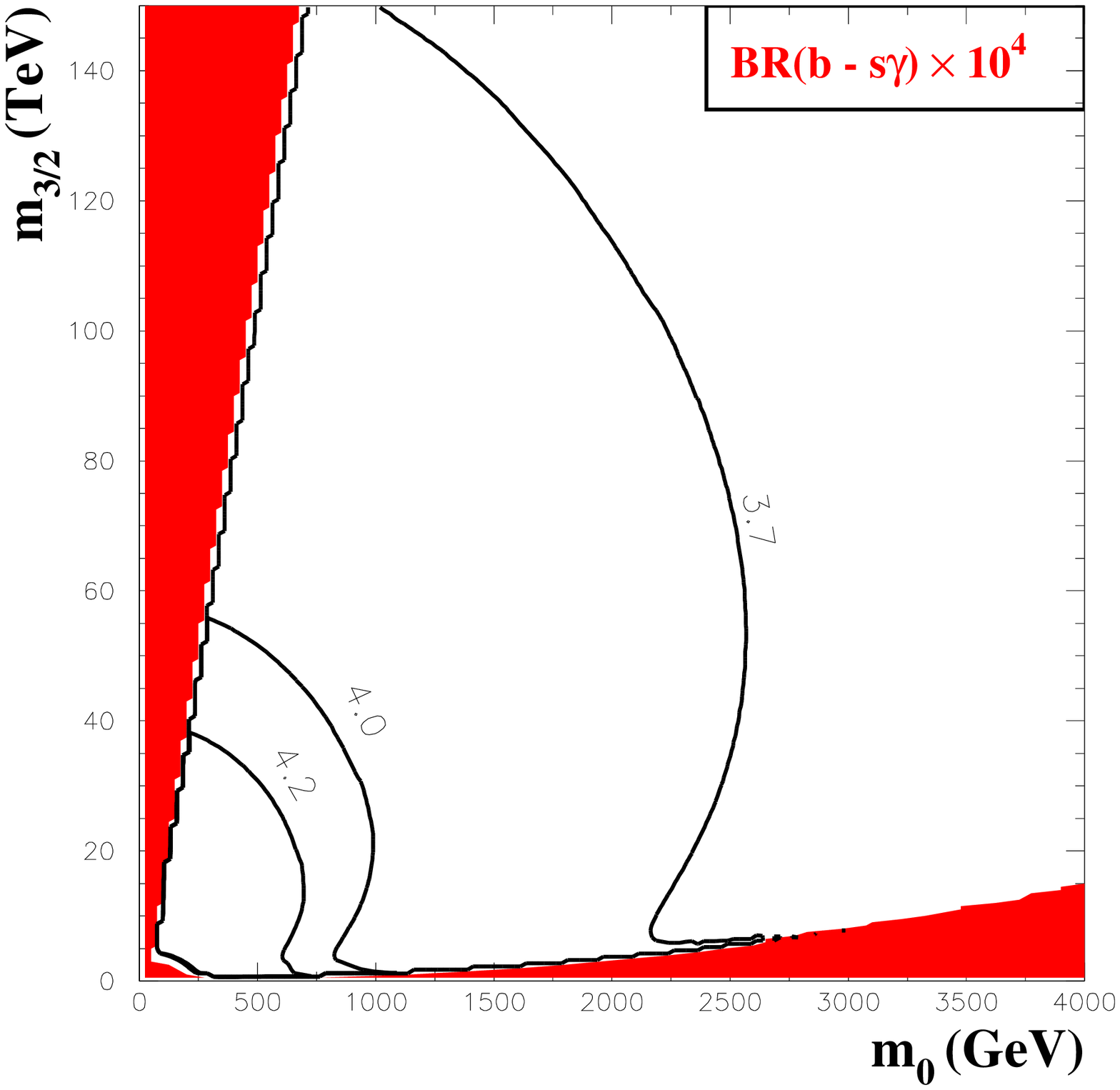}
\includegraphics[width=7.5cm,height=8.0cm]{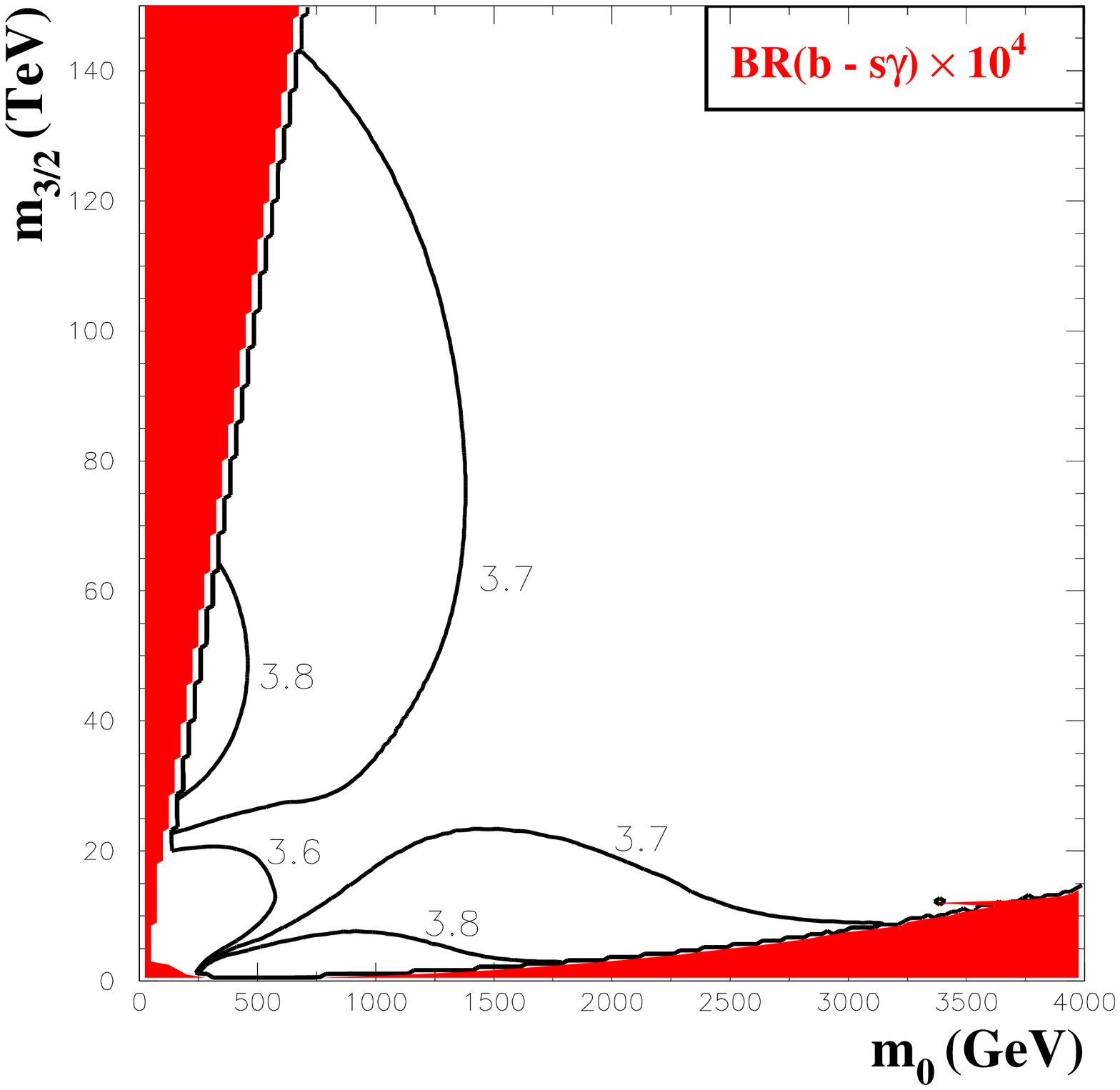}
\caption{The rate of $b \rightarrow s \gamma$ for
$\tan\beta=10$ and $\mu>0$ (left), $\mu<0$ (right panel).}
\label{fig:bsg}
\end{figure}
%%%%%%%%%%%%%%%%%%%%%%%%%%%Figure%%%%%%%%%%%%%%%%%%%%%%%%%%%%%%%%%

The value of the supersymmetric contribution ($\delta a_\mu$) to the
magnetic moment of the muon is shown in Fig.~\ref{fig:g2} for
$\tan\beta=10$.  The most recent data shows a 3.4$\sigma$ disagreement
with the Standard Model (SM) value, having $\delta a_\mu = (27.6 \pm 8.1)
\times 10^{-10}$ ~\cite{a2}.  As is well known the sign of $\delta
a_\mu$ tracks the sign of the $\mu$-parameter, in theories where the
wino and the bino mass are positive. Therefore, clearly the case $\mu
< 0$ is disfavored.

The value of BR($b \rightarrow s \gamma$) is shown in Figure
\ref{fig:bsg} for $\tan\beta=10$.  The current experimental value
reads, ${\rm BR}(b \rightarrow s \gamma)=(3.55 \pm 0.26) \times
10^{-4}$, which is in fine agreement with the Standard Model
expectations.  The supersymmetric contribution to this rare process is
enhanced by $\tan\beta$~\cite{bsg} and in general the rate only
deviates from the standard model value for large $\tan\beta$.
Therefore, the whole region in Fig.~\ref{fig:bsg} lies within the
3$\sigma$ range.  Please note that as the gluino mass is negative in
anomaly mediated supersymmetry one does not find a preference for
$\mu>0$ as in the mSUGRA scenario.

For large values of $\tan\beta$ (about 40) and $\mu>0$ the neutrino
data cannot be fitted for low values of $m_0$.  Moreover, for large
$\tan\beta$ the bound on the lightest neutral Higgs, as well as the
rate of $b \rightarrow s \gamma$, rules out a large corner of the low
$m_0$ and low $m_{3/2}$ parameter space. Furthermore, for $\mu<0$ the
constraints from $g-2$ of the muon excludes all of the parameter-space
at 3$\sigma$ for $\tan\beta=10$. As the supersymmetric contribution to
the magnetic moment of the muon is proportional to $\tan\beta$, the
disagreement with the measured values only becomes worse at large
$\tan\beta$.  Henceforth, we will focus our analysis at a relative low
value of $\tan\beta$, which we fix to be 10, and assume that $\mu$ is
positive.

%%%%%%%%%%%%%%%%%%%%%%%%%%%%%%%%%%%%%%%%%%%%%%%%%%%%%%%%%%%%
\subsection{Neutrino Parameters}

In this section we study a case solution for neutrino physics within
the context of BRpV-mAMSB. We concentrate in the following point of
mAMSB parameter space,
\begin{equation}
m_{3/2}=40\,{\mathrm{TeV}}, 
m_0=500\,{\mathrm{GeV}},
\tan\beta=10, 
\mu>0
\end{equation}
which leads to reasonable values for $B(b\rightarrow s\gamma)$ and
$\delta a_{\mu}$, as can be seen from Figs.~\ref{fig:g2} and
\ref{fig:bsg}. In the spectrum of this model, the Higgs sector is
characterized by a light Higgs mass of $m_h=111.4$ GeV and a
relatively heavy charged Higgs with $m_{H^\pm}=834$ GeV. The LSP is
the first neutralino with $m_{\chi^0_1}=120.85$ GeV, followed by a
nearly degenerate chargino with $m_{\chi^+_1}=120.88$ GeV. The
lightest slepton is the stau with $m_{\tau_1}=458$ GeV, and the
lightest squark is the stop with $m_{t_1}=672$ GeV.

The main effect in collider physics of the presence of BRpV is the 
instability of the neutralino. But in addition, in the neutrino sector we 
gain a mechanism for generating masses for the neutrinos, which in turn 
explain their oscillations. An example solution for neutrino physics, which 
we call benchmark 1, is given by the following BRpV parameters,
\begin{eqnarray}
\epsilon_1=-0.0117\,, &
\epsilon_2=-0.43\,, &
\epsilon_3=-0.246\,{\mathrm{GeV}}
\nonumber\\ 
\Lambda_1=-0.0467\,, &
\Lambda_2= 0.00305\,, &
\Lambda_3= 0.0689\,{\mathrm{GeV}^2} \;.
\label{epslam}
\end{eqnarray}
It predicts the following neutrino observables,
\begin{eqnarray}
&\Delta m_{\rm atm}^2=2.4\times 10^{-3}\,{\mathrm{eV}^2}, 
\Delta m_{\rm sol}^2=8.0\times 10^{-5}\,{\mathrm{eV}^2},
% m_{ee}=0.0044\,{\mathrm{eV}}
\nonumber\\ 
&\tan^2\theta_{\rm atm}=1.27\,, 
\tan^2\theta_{\rm sol}=0.49\,, 
\tan^2\theta_{\rm reac}=0.027\, \;.
\end{eqnarray}
These results are calculated from the full one-loop renormalized
$7\times7$ neutralino-neutrino mass matrix. 

In order to gain some insight into the problem, we study next some 
approximations. It is known that for small BRpV parameters, the 
$3\times3$ effective neutrino mass matrix takes the form,
\begin{equation}
M^\nu_{ij}=A\Lambda_i\Lambda_j
         +B(\Lambda_i\epsilon_j+\Lambda_j\epsilon_i)
         +C\epsilon_i\epsilon_j
\label{mnuij}
\end{equation}
where $A$ receives contributions from tree-level as well as one-loop, and
$B$ and $C$ are one-loop generated. These three parameters depend only 
on MSSM masses and couplings and not on BRpV parameters. All the dependence 
on BRpV is in the $\epsilon_i$ and $\Lambda_i$. From the $7\times7$ mass
matrix in benchmark 1, the corresponding numerical values for the $A$, $B$,
and $C$ parameters of the $3\times3$ effective mass matrix are,
\begin{equation}
A\approx -2.10\,{\mathrm{eV/GeV}^4},\qquad 
B\approx 0.157\,{\mathrm{eV/GeV}^3},\qquad 
C\approx -0.162\,{\mathrm{eV/GeV}^2}
\end{equation}
%

%%%%%%%%%%%%%%%%%%%%%%%%%%%Figure%%%%%%%%%%%%%%%%%%%%%%%%%%%%%%%%%
\begin{figure}[t]
\centering
\includegraphics[height=8.0cm,angle=-90]{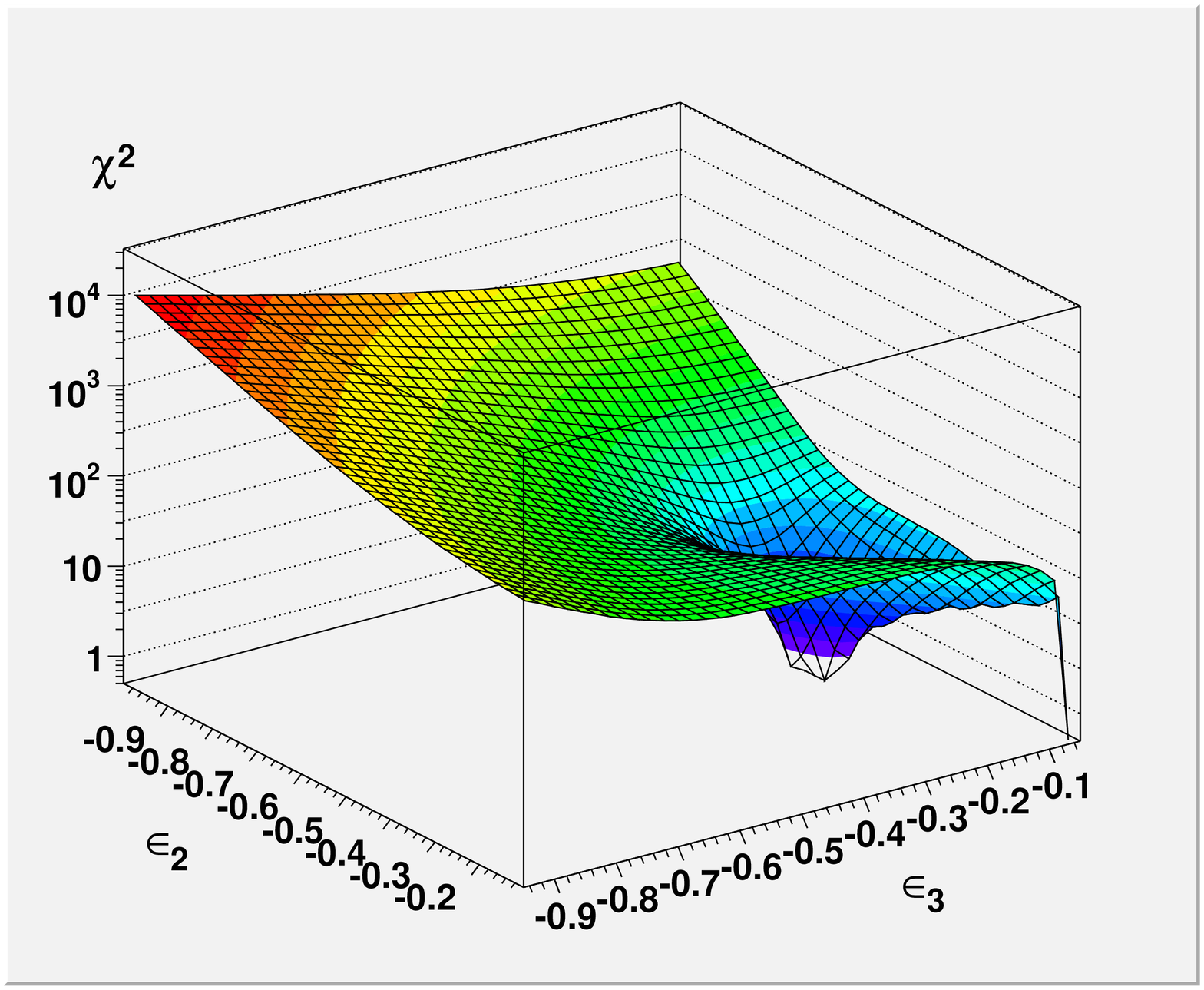}
\includegraphics[height=8.0cm,angle=-90]{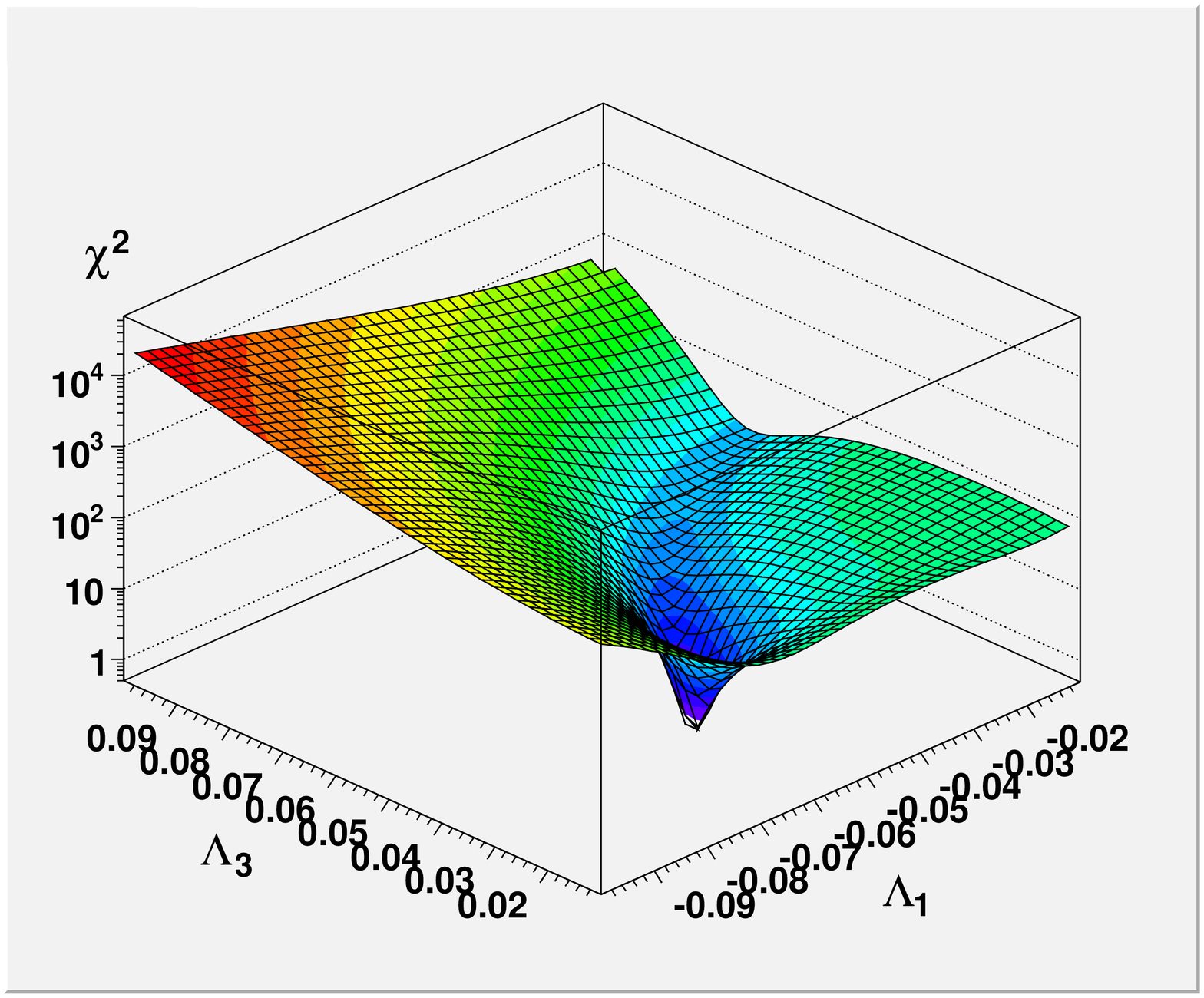}
\caption{$\chi^2$ in the $\epsilon_2$-$\epsilon_3$ plane for the left 
frame and in the $\Lambda_1$-$\Lambda_3$ plane for the right frame.}
\label{fig:X2}
\end{figure}
%%%%%%%%%%%%%%%%%%%%%%%%%%%Figure%%%%%%%%%%%%%%%%%%%%%%%%%%%%%%%%%

The error we make when we use the approximation in eq.~(\ref{mnuij})
can be estimated with a $\chi^2$ evaluated at the input values given
in eq.~(\ref{epslam}), where we defined $\chi^2$ as
\begin{equation}
\chi^2=
\left(\frac{\Delta m^2_{\rm atm}-2.35}{0.95}\right)^2+
\left(\frac{\Delta m^2_{\rm sol}-8.15}{0.95}\right)^2+
\left(\frac{\sin^2\theta_{\rm atm}-0.51}{0.17}\right)^2+
\left(\frac{\sin^2\theta_{\rm sol}-0.305}{0.075}\right)^2
\label{chi2def}
\end{equation}
where the central values and $3\sigma$ deviations were taken from
ref.~\cite{Maltoni:2004ei}. The atmospheric mass difference is given 
in $10^{-3}\,{\mathrm{eV}}^2$, and the solar mass difference in
$10^{-5}\,{\mathrm{eV}}^2$. Both are defined to be positive.
The neutrino observables calculated with the effective $3\times3$ mass
matrix give the value $\chi^2=5.7$, most of it coming from the solar 
angle, indicating the kind of error we make when we use it.

For illustrative purposes, we find now the least modified values of
BRpV parameters that give a good solution for neutrino observables 
calculated using the diagonalization of the effective $3\times3$ neutrino 
mass matrix in eq.~(\ref{mnuij}). We call it benchmark 1':
\begin{eqnarray}
\epsilon_1=-0.0117\,, &
\epsilon_2=-0.50\,, &
\epsilon_3=-0.16\,{\mathrm{GeV}} \;,
\nonumber\\ 
\Lambda_1=-0.064\,, &
\Lambda_2= 0.00305\,, &
\Lambda_3= 0.033\,{\mathrm{GeV}^2} \, \;.
\label{epslamp}
\end{eqnarray}
The difference between this benchmark 1' and the one in eq.~(\ref{epslam})
indicates us how erred would be the determination of BRpV parameters if
we do not use the full $7\times7$ mass matrix in the calculation of the 
neutrino observables.

In Fig.~\ref{fig:X2} we plot $\chi^2$, which 
measures the deviation of a given model prediction from the experimental 
measurements. We calculate $\chi^2$ using the $3\times3$ effective mass
matrix, and we use benchmark 1'. In the left frame we vary $\epsilon_2$
and $\epsilon_3$, keeping all the other parameters constant as indicated
by benchmark 1'. In the right frame we vary $\Lambda_1$ and $\Lambda_3$.
By construction, the minimum appear at the values defined by benchmark 1'.
The fact that small deviations on the BRpV parameters produce very large
values of $\chi^2$ indicate us how sensible are the neutrino observables to
them. 

To explore in more detail the dependence of each neutrino observable 
on BRpV parameters we define the quantities $\chi^2_i$, $i=1,4$ as the 
$\chi^2$ calculated using only the $i$-th term in eq.~(\ref{chi2def}). 
As in Fig.~\ref{fig:X2}, the $\chi^2_i$ are calculated using the 
$3\times3$ effective neutrino mass matrix approximation.
%%%%%%%%%%%%%%%%%%%%%%%%%%%Figure%%%%%%%%%%%%%%%%%%%%%%%%%%%%%%%%%
\begin{figure}[t]
\centering
\includegraphics[height=8.0cm,angle=-90]{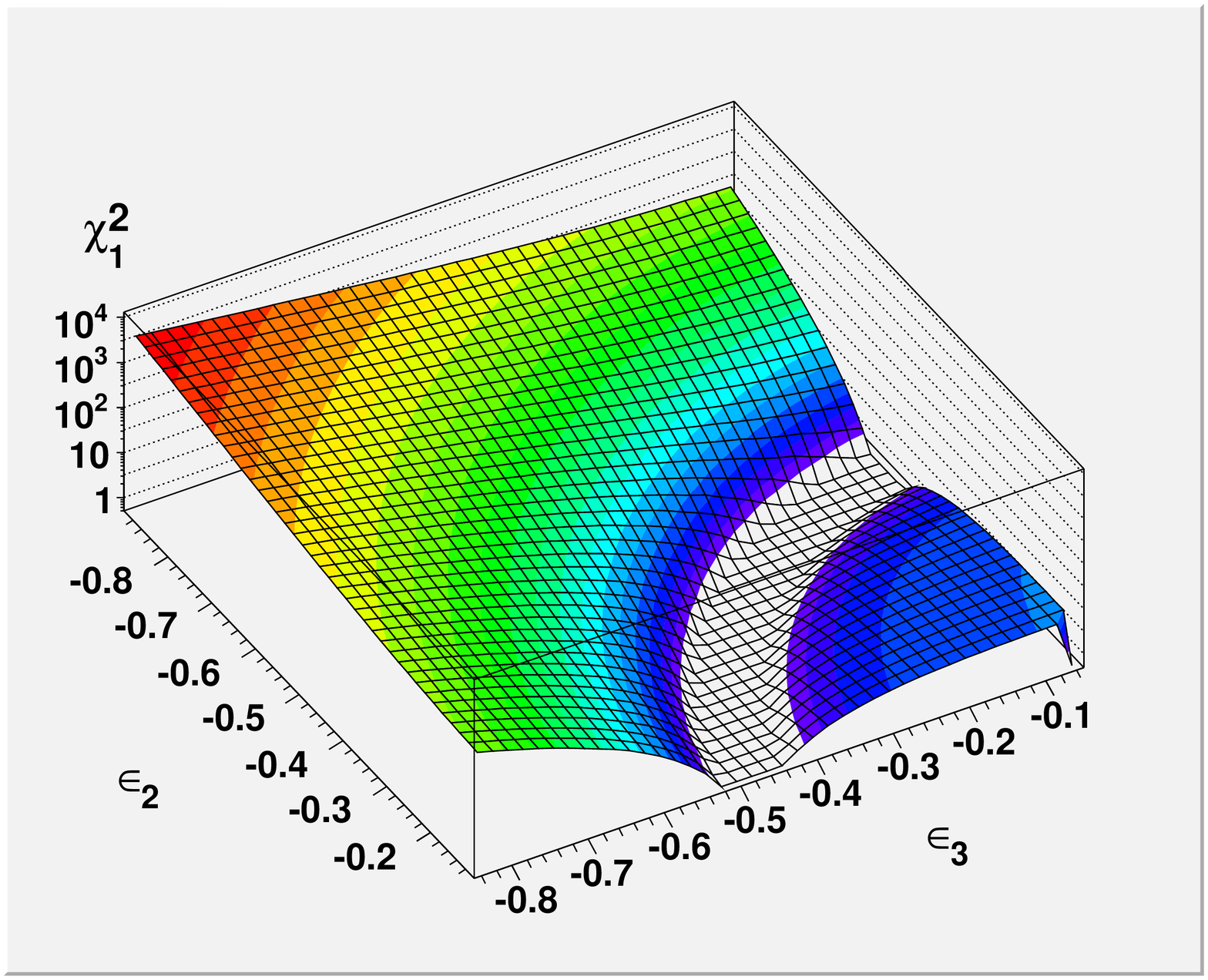}
\includegraphics[height=8.0cm,angle=-90]{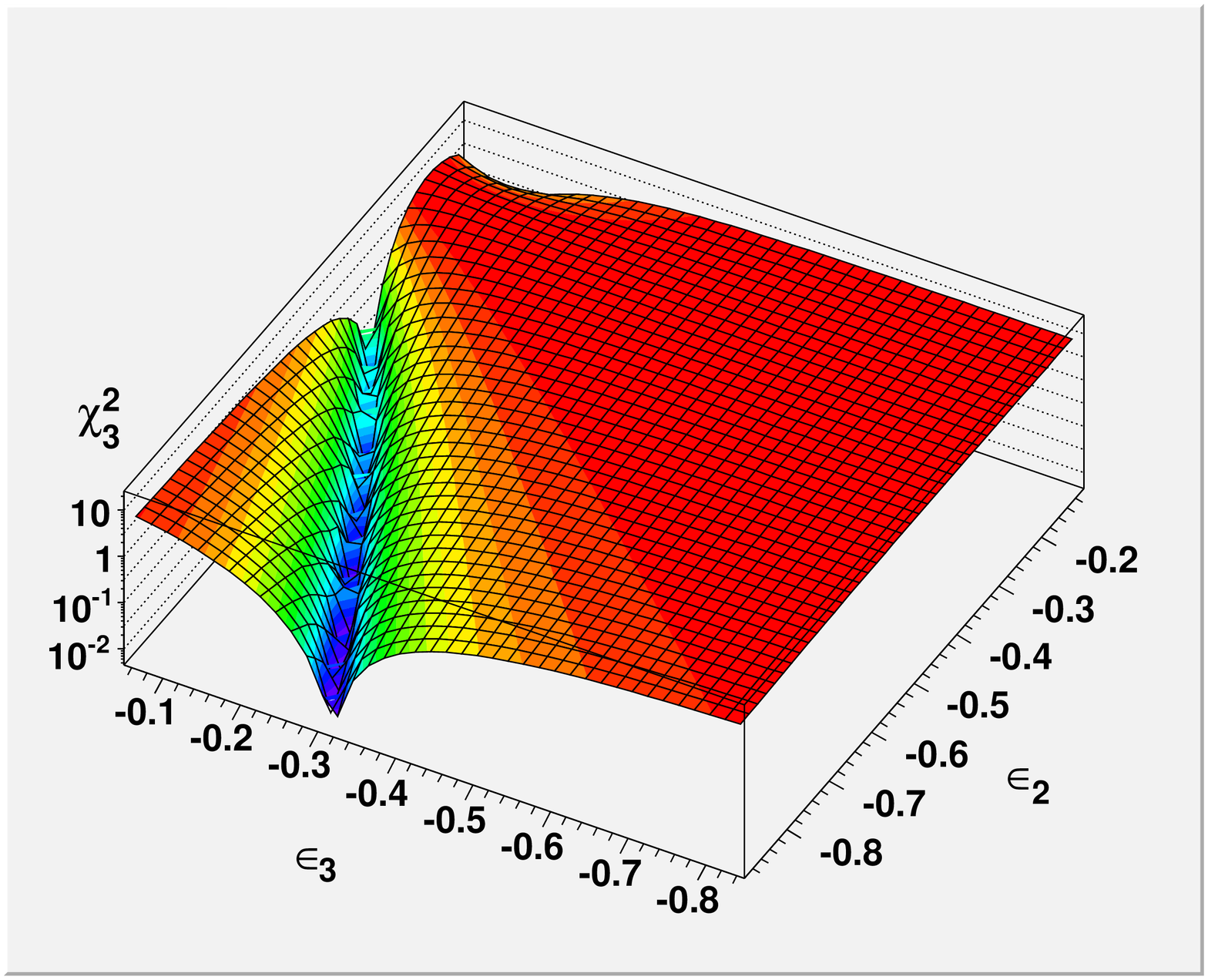}
\caption{Partial $\chi^2$ in the $\epsilon_2$-$\epsilon_3$ plane:
atmospheric mass for the left frame, and atmospheric angle for the 
right frame.}
\label{fig:X2atm}
\end{figure}
%%%%%%%%%%%%%%%%%%%%%%%%%%%Figure%%%%%%%%%%%%%%%%%%%%%%%%%%%%%%%%%
We start with atmospheric parameters in Fig.~\ref{fig:X2atm}. In the 
left frame we have $\chi^2_1$,
\begin{equation}
\chi^2_1=\left(\frac{\Delta m^2_{\rm atm}-2.35}{0.95}\right)^2
\end{equation}
which is no more than the first term in 
eq.~(\ref{chi2def}), associated with the atmospheric mass squared 
difference, plotted as a function of $\epsilon_2$ and $\epsilon_3$.
In the right frame we have the solar angle represented by $\chi^2_3$,
as a function of the same parameters. The behaviour of $\chi^2_i$
can be easily understood with some approximations which we develop next.

In normal circumstances the $A$-term dominates over the other two in
eq.~(\ref{mnuij}) because it receives contributions at tree-level. 
Nevertheless, depending on the relative values of the $\epsilon_i$ and 
$\Lambda_i$ parameters, it is possible for the $C$-term to dominate the 
neutrino mass matrix. This is the case with the example we 
are studying. We have obtained approximated solutions for eigenvalues and 
eigenvectors of the $3\times3$ effective neutrino mass matrix when the 
$C$-term is much larger than the $A$ and $B$ terms. In this case, the 
neutrino masses are,
\begin{eqnarray}
m_3&=&C|\vec\epsilon|^2
+2B(\vec\epsilon\cdot\vec\Lambda)
+A\frac{(\vec\epsilon\cdot\vec\Lambda)^2}{|\vec\epsilon|^2}
\nonumber\\
m_2&=&
A\frac{|\vec\epsilon\times(\vec\Lambda\times\vec\epsilon)|^2}
{|\vec\epsilon|^4}
\label{miapp}
\end{eqnarray}
up to terms of second order, while the lightest neutrino has exactly 
$m_1=0$ when the mass matrix has the form in eq.~(\ref{mnuij}). The 
eigenvectors are given by the following expressions,
\begin{eqnarray}
%\textcolor[rgb]{0.9,0.4,0}
{\vec v_3}&=&
\frac{\vec\epsilon}{|\vec\epsilon|}
+\left[B+A\frac{(\vec\epsilon\cdot\vec\Lambda)}{|\vec\epsilon|^2}
\right]
\frac{\vec\epsilon\times(\vec\Lambda\times\vec\epsilon)}
{C|\vec\epsilon|^3}
\nonumber\\
%\textcolor[rgb]{0.9,0.4,0}
{\vec v_2}&=&
\frac{\vec\epsilon\times(\vec\Lambda\times\vec\epsilon)}
{|\vec\epsilon\times(\vec\Lambda\times\vec\epsilon)|}
-\left[B+A\frac{\vec\epsilon\cdot\vec\Lambda}{|\vec\epsilon|^2}
\right]\frac{|\vec\epsilon\times(\vec\Lambda\times\vec\epsilon)|}
{C|\vec\epsilon|^4}\vec\epsilon
\label{viapp}
\\
%\textcolor[rgb]{0.9,0.4,0}
{\vec v_1}&=&
\frac{\vec\Lambda\times\vec\epsilon}
{|\vec\Lambda\times\vec\epsilon|}
\nonumber
\end{eqnarray}
also up to terms of second order. Note that the eigenvectors are 
orthogonal and normalized up to the order we are working. The matrix 
$U_{PMNS}$ is formed with the eigenvectors in its columns.

%%%%%%%%%%%%%%%%%%%%%%%%%%%Figure%%%%%%%%%%%%%%%%%%%%%%%%%%%%%%%%%
\begin{figure}[t]
\centering
\includegraphics[height=8.0cm,angle=-90]{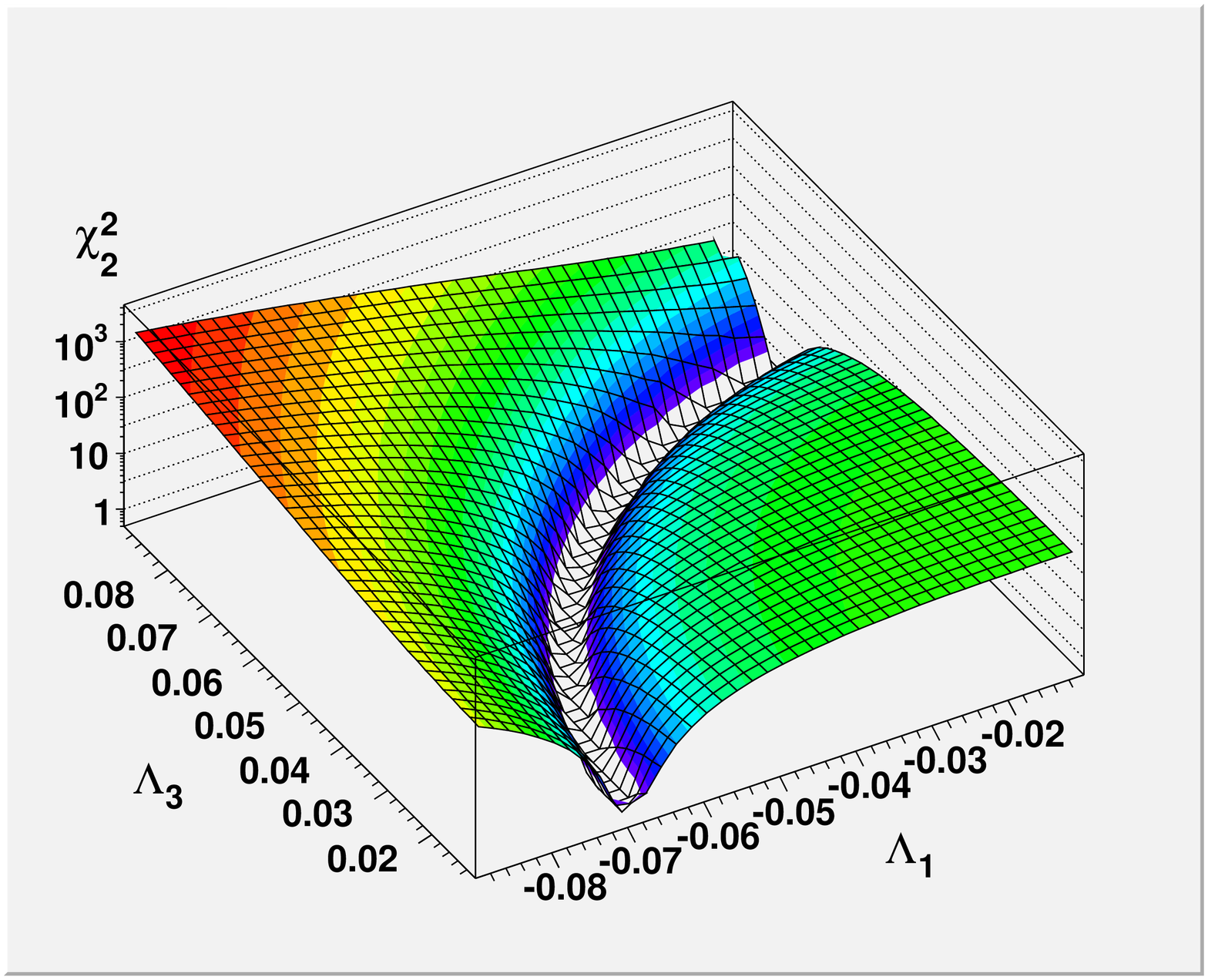}
\includegraphics[height=8.0cm,angle=-90]{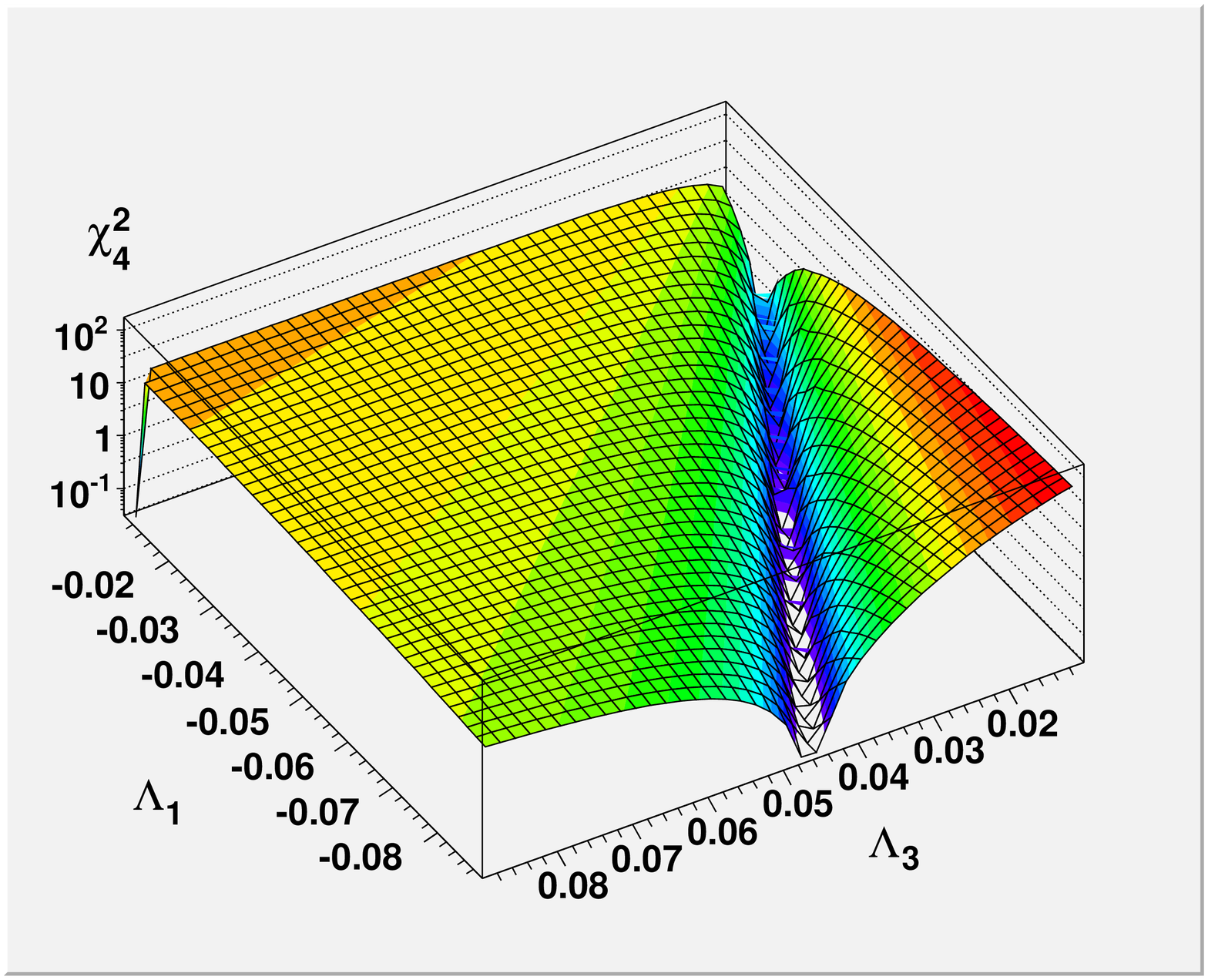}
\caption{Partial $\chi^2$ in the $\Lambda_1$-$\Lambda_3$ plane:
solar mass for the left frame, and solar angle for the 
right frame.}
\label{fig:X2sol}
\end{figure}
%%%%%%%%%%%%%%%%%%%%%%%%%%%Figure%%%%%%%%%%%%%%%%%%%%%%%%%%%%%%%%%
Using these approximated expressions, and neglecting $\Lambda_2$ and
$\epsilon_1$, we find,
\begin{eqnarray}
\Delta m_{\rm atm}^2=m_3^2-m_2^2 &\approx& C^2(\epsilon_2^2+\epsilon_3^2)^2
\nonumber\\
\Delta m_{\rm sol}^2=m_2^2-m_1^2 &\approx& A^2\left[
\Lambda_1^2+\frac{\Lambda_3^2}{1+(\epsilon_3/\epsilon_2)^2}\right]^2
\nonumber\\
\tan^2\theta_{\rm atm}=\left(\frac{v_{3,2}}{v_{3,3}}\right)^2 &\approx& 
\left(\frac{\epsilon_2}{\epsilon_3}\right)^2
\label{approxim}\\
\tan^2\theta_{\rm sol}=\left(\frac{v_{2,1}}{v_{3,1}}\right)^2 &\approx& 
\frac{\Lambda_1^2}{\Lambda_3^2}\left[1+
\left(\frac{\epsilon_3}{\epsilon_2}\right)^2\right]
\nonumber
\end{eqnarray}
With these approximations for the neutrino observables, we can easily
understand the different figures. The atmospheric mass squared difference
$\Delta m^2_{\rm atm}$  in eq.~(\ref{approxim}) indicates that constant values 
of this observable are obtained at circumferences in the 
$\epsilon_2$-$\epsilon_3$ plane, which is exactly what we see in the left 
frame of Fig.~\ref{fig:X2atm}. At the same time, constant values of the 
atmospheric angle are obtained at straight lines, which is confirmed in 
the right frame of Fig.~\ref{fig:X2atm}. From eq.~(\ref{approxim}) we 
also see that the dependence on $\Lambda_i$ of the atmospheric parameters 
is weak, and we do not plot it explicitly.

In Fig.~\ref{fig:X2sol} we concentrate on the solar neutrino parameters, 
and we study them as a function of $\Lambda_1$ and $\Lambda_3$. From 
eq.~(\ref{approxim}) we see that the solar mass squared difference has a 
constant value at ellipses in the $\Lambda_1$-$\Lambda_3$ plane, and they 
can be seen in the left frame of Fig.~\ref{fig:X2sol}, with the 
eccentricity of the ellipse depending on the ratio $\epsilon_3/\epsilon_2$.
Similarly, in eq.~(\ref{approxim}) we learn that constant values of the 
solar angle are obtained at straight lines in the $\Lambda_1$-$\Lambda_3$ 
plane, and they can be seen in the right frame of Fig.~\ref{fig:X2sol}.
The dependence of the solar observables on $\epsilon_i$ is not weak, but
we do not show it here.

%%%%%%%%%%%%%%%%%%%%%%%%%%%%%%%%%%%%%%%%%%%%%%%%%%%%%%%%%%%%%%%%%%%%%%
\subsection{R--parity violating decays of SUSY particles}

The R--parity violating interactions are rather feeble since they are
related to neutrino physics. Therefore, the R--parity violating
effects are expected to be small except in processes suppressed in the
$R$ conserving scenario.
In fact, the pair production of SUSY particles is nearly the same as
in the case of conserved R parity and single production of SUSY
particles is strongly suppressed. 
%however, the decay of the lightest supersymmetric particles is
%significantly altered.
The main manifestation of R--parity violation is the fact that 
the lightest supersymmetric particles decays. 

We evaluated all possible R--parity conserving as well as R--parity
violating decays for all particles.  Notwithstanding, with the
exception of $\chi^0_1$ and $\chi^{+}_1$ the R--parity violating
channels are strongly suppressed as they are proportional to the
ratios $|\vec \epsilon_i|^2 / |\mu|^2 \simeq 10^{-6}$ and $|\vec
\Lambda|^2 / |\det(m_{\chi^0})| \simeq 10^{-8}$ and, thus, do not play
any role in our analysis.  Due to the near degeneracy of the lightest
neutralino and chargino the decay of the latter through R--parity
violating couplings is significant (often dominant) and we evaluate
all RPV and RPC decay channels.  The decay of $\chi^0_1$ and
$\chi_1^+$ are calculated using the two-body decay whenever possible.
Thus, for masses above the $W$-mass we compute the decays $\chi^0_1
\rightarrow \ell_i^\pm W^\pm$, whereas below $M_W$ we compute the full
three--body decay into all possible final states.  In the three--body
decays, we include all intermediates states, such as neutral scalars
and pseudo--scalars which can be important in some regions of
parameter space~\cite{porod1}.  However, the effects from the scalar
intermediate states, except the Standard Model like Higgs, are less
import in our AMSB model, as the scalars are fairly heavy.

%%%%%%%%%%%%%%%%%%%%%%%%%%%Figure%%%%%%%%%%%%%%%%%%%%%%%%%%%%%%%%%
\begin{figure}[t]
\centering
\includegraphics[width=8.cm,height=9.0cm]{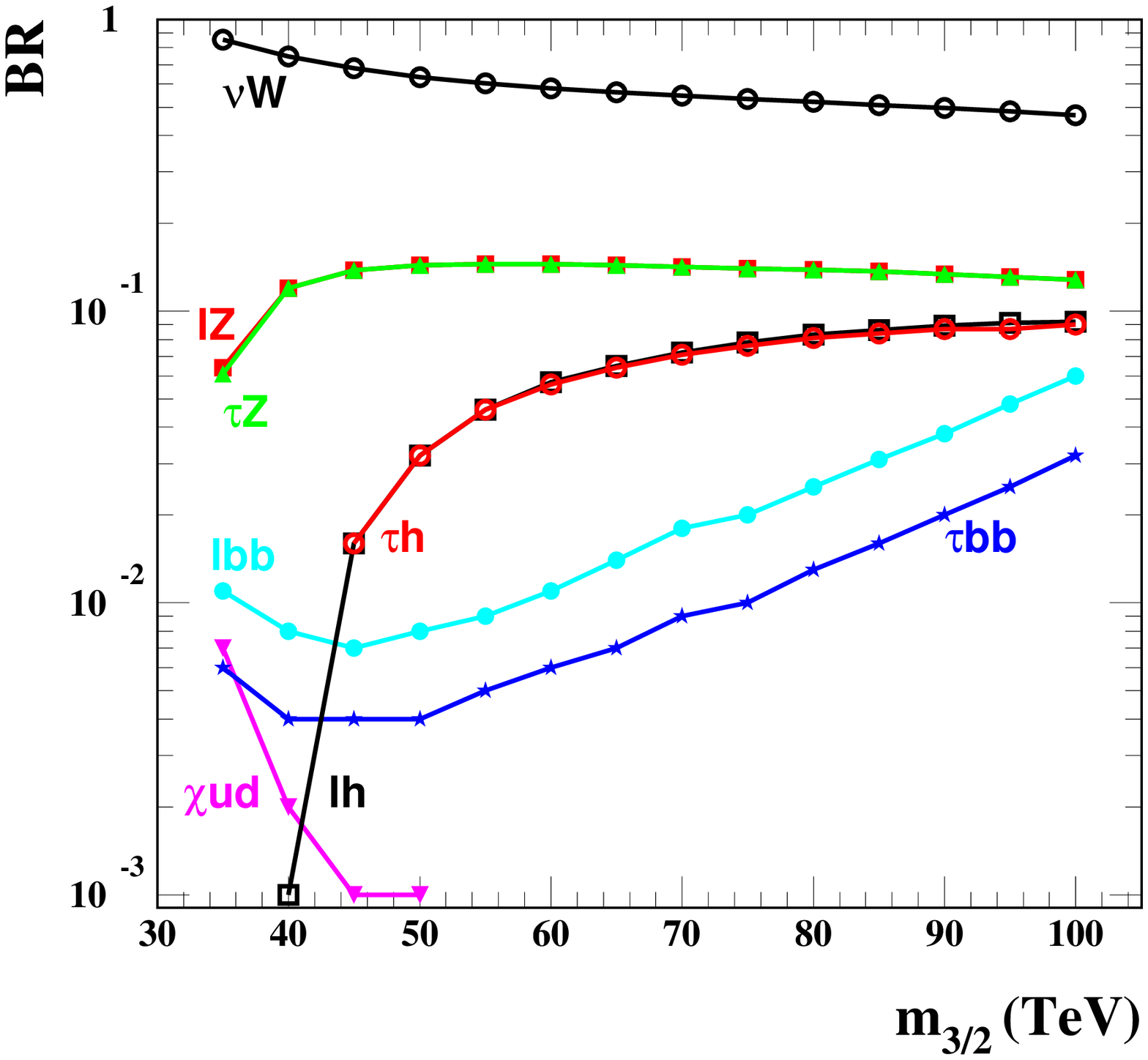}
\includegraphics[width=8.cm,height=9.0cm]{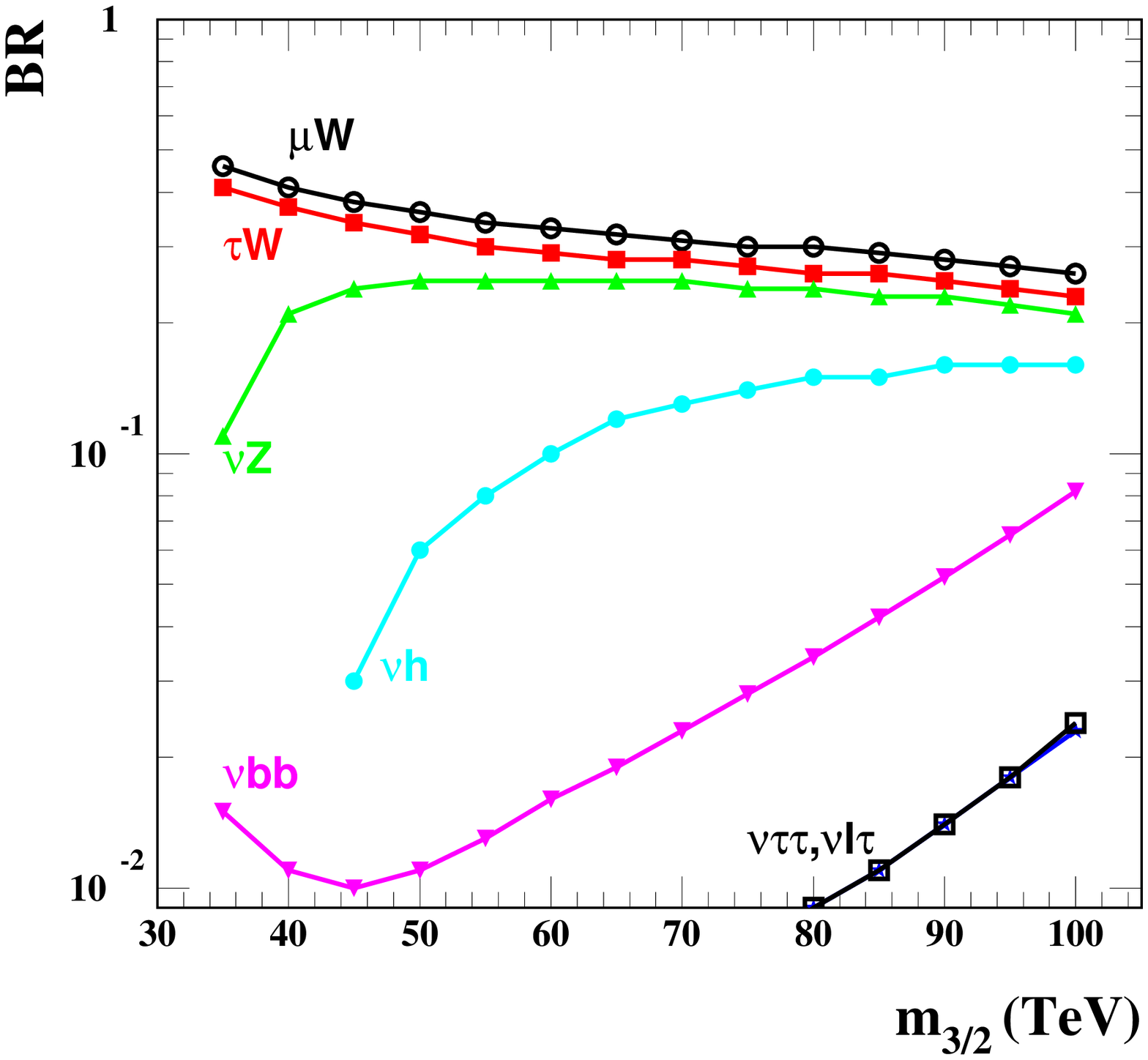}
\caption{The branching ratios of the lightest chargino (left) and
the lightest neutralino (right) as a function of $m_{3/2}$ for
$m_0=800$ GeV, $\tan\beta=10$, and $\mu>0$.  We use the symbol
l for an electron or a muon. By $q$ we mean any quark not being
the bottom quark, and by the $ud$ combinations we mean any set
of u-type quarks and d-type quarks. In general we sum
over all states in the decay channel. 
}
\label{fig:br}
\end{figure}
%%%%%%%%%%%%%%%%%%%%%%%%%%%%%%%%%%%%%%%%%%%%%%%%%%%%%%%%%%%%

The decay of the lightest neutralino in our model takes place through
$W,Z$, neutral and charged scalar and squark states. The decays of the
neutralino can be classified as leptonic ($\nu \ell \ell$),
semileptonic ($\ell q \bar q'$, $\nu q \bar{q}$) as well as invisible
($\nu\nu\nu$).  The possible BRpV chargino decay channels, induced
through the same intermediate states as for the neutralino, are $\nu
q' \bar{q}$, $\ell q \bar{q}$, $\ell \ell \ell$ and $\ell \nu \nu$.
Moreover, R--parity conserving chargino decays exist with the most
important RPC channels being $\chi_1^+ \rightarrow \chi_1^0 \pi^+$ and
$\chi_1^+ \rightarrow \chi_1^0 e^+ \nu$.  The mass difference $\Delta
m_{\chi}$, is mainly dependent on the value of $\tan\beta$ and
$m_{3/2}$, and is so small that it suppresses the RPC channel
substantially.  For this reason the RPC decays will have the largest
branching ratio for small $m_{3/2}$ and large $\tan\beta$, since in
this case $\Delta m_\chi$ is the largest.

To understand better the decay of these particles we plot their
branching ratios in Figure~\ref{fig:br}.  The chargino decay channels
with a branching ratio above $10^{-3}$ are plotted and for the
neutralino we plot all channels above $10^{-2}$. We have
differentiated between bottom quarks (denoted b) and other quarks
(denoted q). Also, we distinguish between a muon or an electron
(commonly denoted by $l$) and a tau.
We note, that as has already been observed~\cite{rpvcol,porod1} there
is an important connection between the neutralino decay and the
neutrino parameters. In particular, due to the large atmospheric
mixing angle, an almost equal number of muons and taus is expected to
be produced along with a W. This can clearly
be seen from Fig.~\ref{fig:br}, as the W-mediated channels exhibit
this property. We can also learn from this figure that for heavier
neutralinos, {\em i.e.} larger $m_{3/2}$, the most important decay
channels are $\mu W$, $\tau W$, and $\nu Z$ with the $\nu h$ mode
having a sizeable contribution. The three--body decays grow with the
increase of $m_{3/2}$, becoming important at high $m_{3/2}$ values.

The chargino branching ratios also exhibit a number of near
equalities. For instance, from the right panel of Fig.~\ref{fig:br} we
can see that BR$(\chi^+_1 \rightarrow \ell Z) ~\simeq$ BR$(\chi^+_1
\rightarrow \tau Z)$ as well as BR$(\chi^+_1 \rightarrow \ell h)
~\simeq$ BR$(\chi^+_1 \rightarrow \tau h)$.  Indeed, these equalities
are found for all values of $m_{3/2}$, being a consequence of the fact
that the same R-parity breaking parameters responsible for the SUSY
decays govern also neutrino physics \cite{Hirsch:2003fe}.
Furthermore, the chargino decays predominantly into $\nu W$ for all
values of $m_{3/2}$ and the other important decays are $\ell Z$, $\tau
Z$, $\ell h$ and $\tau h$.  The mass difference $\Delta m_\chi$ for
$\tan\beta=10$ is at most 300 MeV and as can be seen from
Fig.~\ref{fig:br}, the chargino RPC branching ratios are very small
and can be neglected in the analysis.

%%%%%%%%%%%%%%%%%%%%Section%%%%%%%%%%%%%%%%%%%%%%%%%%%%%%%%%
\section{Collider signals}\label{signals}

We will analyze the LHC discovery potentials for BRpV-mAMSB in various
channels, that is, we study a myriad of channels, ranging from
jets+missing energy to multilepton channels in addition to the displaced
vertices signal. Throughout this paper we use SPheno~\cite{spheno} to
generate the particle spectrum and decays which are tabulated in the
SLHA format~\cite{Skands:2003cj}.  The signal and background
generation was carried out with PYTHIA~\cite{pythia} version 6.409
adopting the CTEQ5L parton distribution function~\cite{Lai:1999wy}.

%%%%%%%%%%%%%%%%%%%%%%%%%%%%%%%%%%%%%%%%%%%%%%%%%%
\subsection{Canonical SUSY final state topologies at
  LHC}\label{toplhc}

We considered several canonical supersymmetry signals for the LHC,
following what has been presented in Refs.~\cite{Baer:2000bs} and
\cite{deCampos:2007bn}:

\begin{enumerate}

\item {\em Inclusive jets and missing transverse momentum} {\bf(IN)}:
  in this class of events we include all events that present jets and
  missing $\sla{p}_T$. In R--parity conserving scenarios this channel
  is one of the main search modes~\cite{Baer:2000bs}.

\item {\em Zero lepton, jets and missing transverse momentum}
  $\mathbf{(0\ell)}$: the events in this class present jets and
  missing $\sla{p}_T$ without isolated leptons ($e^\pm \,,\,
  \mu^\pm$);

\item {\em One lepton, jets and missing transverse momentum}
  $\mathbf{(1\ell)}$: here we consider only events presenting jets and
  missing $\sla{p}_T$ accompanied by just one isolated lepton;

\item {\em Opposite sign lepton pair, jets and missing transverse
    momentum} {\bf (OS)}: the events in this class contain jets,
  missing $\sla{p}_T$, and two isolated leptons of opposite charges;

\item {\em Same sign lepton pair, jets and missing transverse
    momentum} {\bf (SS)}: here we consider only events presenting jets
  and missing $\sla{p}_T$ accompanied by two isolated leptons of the
  same charge;

\item {\em Multileptons, jets and missing transverse momentum}
  $\mathbf{(M\ell)}$: we classify in this class the events exhibiting
  jets, missing $\sla{p}_T$ accompanied by three or more isolated charged
  leptons.

\end{enumerate}

In our analysis we defined jets through the subroutine PYCELL of
PYTHIA with a cone size of $\Delta R = 0.7$. Charged leptons ($e^\pm$
or $\mu^\pm$) were considered isolated if the energy deposit in a cone
of $\Delta R < 0.3$ is smaller than 5 GeV.  Furthermore, we smeared
the energies, but not directions, of all final state particles with a
Gaussian error given by $\Delta E/E = 0.7/\sqrt{E}$ ($E$ in GeV) for
hadrons and $\Delta E /E = 0.15/\sqrt{E}$ for charged leptons.

We perform our event selection along the same lines of
\cite{Baer:2000bs}.  Initially, we applied the following acceptance
cuts:

\medskip

\noindent{\bf C1} We required at least {\bf two} jets in the event with
\begin{equation}
  p^j_T  >  50 \hbox{ GeV} \;\;\; \hbox{and} \;\;\; |\eta_j| < 3 \; .
\end{equation}

\noindent{\bf C2} The transverse sphericity of the event must exceed
0.2, {\em i.e.}
\begin{equation}
  S_T > 0.2 \; .
\end{equation}
This requirement reduces efficiently the large background due to  the
production of two jets in the SM.

\medskip

\noindent{\bf C3} A potential source of missing transverse momentum in
the background is the mismeasurement of jets. Therefore, we imposed
the following cut in the azimuthal angle between the jets and the
missing momentum ($\Delta \varphi$) to reduce the background
\begin{equation}
  \frac{\pi}{6} \le \Delta \varphi \le \frac{\pi}{2} \; .
\end{equation}

\medskip  %aqui

After applying the acceptance cuts {\bf C}1--{\bf C}3 we impose
further cuts to each final state topology.  In order to achieve some
optimization of the cuts in different regions of the parameter space
we used floating cuts $E_T^c$ which can take the values 200, 300, 400
and 500 GeV. Considering that our main goal is to evaluate the impact
or R--parity violation interactions on the searches for SUSY we have
not tried to further improve our cuts.  Given one value of $E_T^c$, we
further required for all topologies that
\begin{equation}
 p_T^{1,2} > E_T^c  \;\;\; \hbox{and} \;\;\;
 \sla{p}_T > E_T^c \; .
\label{cuts:1}
\end{equation}
where $ p_T^{1,2}$ stand for the transverse momenta of the two hardest
jets. These are the only additional cuts applied on the {\bf IN}
topology.

For the events in the $0\ell$ class we veto the presence of isolated
leptons with
\begin{equation}
   p_T^\ell > 10 \hbox{ GeV} \;\;\; \hbox{and} \;\;\;  |\eta_\ell| <
   2.5 \; .
\label{cuts:2}
\end{equation}

In addition to comply with (\ref{cuts:1}), $1\ell$ events must present
only one isolated lepton satisfying
\begin{equation}
  p_T^\ell > 20\hbox{ GeV} \;\;\; \hbox{and} \;\;\;  |\eta_\ell| < 2.5
  \; .
\end{equation}
Since the SM production of $W$'s is a potentially large background 
we also imposed that the transverse mass cut
\begin{equation}
  m_T(\sla{p}_T, p^\ell_T) > 100 \hbox{ GeV} \; .
\end{equation}

For the {\bf OS} ({\bf SS}) signal we required the presence of two
isolated charged leptons with opposite (same) charge after imposing
(\ref{cuts:1}). The hardest isolated lepton must have
\begin{equation}
p_T^\ell > 20 \hbox{ GeV} \;\;\; \hbox{and} \;\;\;  |\eta_\ell| < 2.5
\; ,
\label{cuts:3}
\end{equation}
while the second lepton must satisfy (\ref{cuts:2}).
Multilepton events ($M\ell$) must pass the cuts (\ref{cuts:1}) and
exhibit three or more isolated leptons with the hardest one satisfying
(\ref{cuts:3}) and the other leptons complying (\ref{cuts:2}).

%%%%%%%%%%%%%%%%%%%%%%%%%%%%%%%%%%%%%%%%%%%%%
%\subsubsection{Backgrounds}
%%%%%%%%%%%%%%%%%%%%%%%%%%%%%%%%%%%%%%%%%%%%%

The most important SM backgrounds to the canonical SUSY searches are
\begin{itemize}
  
\item the process with highest cross section at small $E_T^c$ is the
  QCD production $pp \to jj X$ where $j$ denotes a jet;

\item $t \bar{t}$ production that contributes to many final state
  topologies due to its decay into $WWbb$;

\item associated production of weak bosons and jets which we denote by
  $Wj$ and $Zj$;

\item double weak boson production $VV$ with $V=Z$ or $W$;

\item production of a single top quark. We did not consider the
  gluon--$W$ contribution to this reaction because it is not included in
  PYTHIA.

\end{itemize}

\begin{table}
\begin{center}
\begin{tabular}{||l|l|l|l|l|l||}
\hline
 $E_T^c$/Background & {\bf IN}  & $\mathbf{1\ell}$ & {\bf OS} & {\bf SS} 
 & $\mathbf{M\ell}$
\\ \hline   
 200 GeV  & 261.      & 178.             & 5.4      & 0.40    & 1.3    
\\ \hline                                                       
 300 GeV  & 24.       & 17.              & 0.32     & 0.013   & 0.001  
\\ \hline               
 400 GeV  & 4.0       & 3.1              & 0.015    & 0.      & 0.     
\\ \hline               
 500 GeV  & 0.88      & 0.63             & 0.003    & 0.      & 0.     
\\
\hline \hline
\end{tabular}
\end{center}
\label{tab:xsec}
\caption{Total background cross section in fb as a function of $E_T^c$
for the channels considered here.}
\end{table}

We present in Table~I %\ref{tab:xsec}
the total background cross section after cuts for the final state
topologies that we analyzed. The main source of background is
$t\bar{t}$ production for all the process with $Wj$ and $Zj$ also
contributing to the {\bf IN} background. Moreover, $Wj$ also plays an
important role in the $\mathbf{1\ell}$ topology.  For additional
information on the backgrounds see Ref.~\cite{deCampos:2007bn}.

Comparing our results for the SM backgrounds with the ones in
Ref.~\cite{Baer:2000bs} we can see that they agree within a factor of
2.  This is indeed expected since PYTHIA and ISAJET have different
choices for the event generation, specially in the hadronization
procedure.  This does not pose a serious problem for the LHC
experiment since the backgrounds can be obtained from actual data.

%%%%%%%%%%%%%%%%%%%%%%%%%%%%%%%%%%%%%%%%%%%%%%%%%%%%%%%%%%%%%%%%%%%%%%
\subsection{Displaced Vertices at LHC}

In our BRpV--mAMSB model both $\chi_1^0$ and $\chi_1^+$ can travel
macroscopic distances before decaying.  Consequently, the long
lifetimes of these particles can give rise to a further striking
signal, that is, the existence of displaced vertices in the events.
However, the decay must be confined within the inner parts of the
detector in order for it to be fully reconstructible.

\begin{figure}[thpb]
\centering
\includegraphics[width=10cm,height=8.0cm]{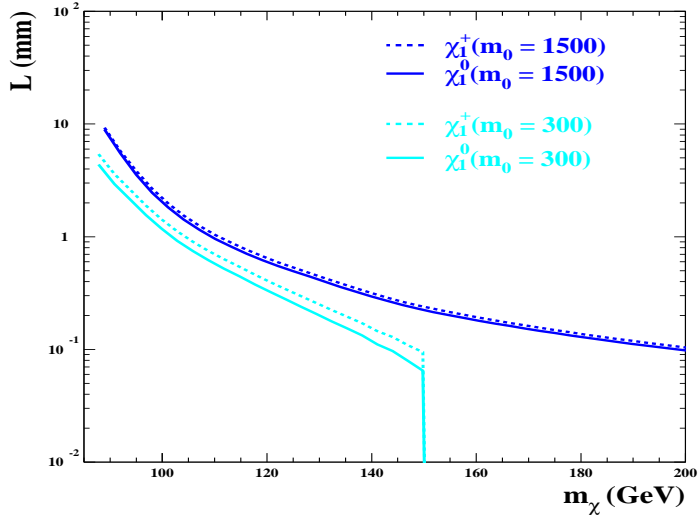}
\caption{The decay length in the rest frame of the lightest neutralino
  and the lightest chargino as a function of the mass for
  $\tan\beta=10$ and $\mu>0$.}
\label{fig:dl_10p}
\end{figure}

The decay lengths in the rest frame of $\chi^0_1$ and $\chi^+_1$ are
depicted in Fig.~\ref{fig:dl_10p} as a function of the neutralino
(chargino) mass for different choices of parameters.  We can see from
this figure that these SUSY particles typically decay within the inner
detector, even when taking into account the time dilation
$\gamma$-factor.  Due to the fact that the sparticles production is
completely dominated by R--parity conserving interactions, there will
almost always be two long lived sparticles in the reaction chain that
may lead to a displaced vertex, either $\chi_1^0$ or $\chi_1^+$. Thus,
our analysis will require the presence of two displaced vertices in
the event.  Notwithstanding, there is a small corner of the parameter
space where $\chi_1^0$ and $\chi_1^+$ decay very fast as we can see in
the bottom curves of Fig.~\ref{fig:dl_10p}. This region is
characterized by the LSP being the stau, and consequently, $\chi_1^0$
and $\chi_1^+$ fast decay is due to R--parity conserving interactions.
Moreover, in this region of the parameter space the stau decay is too
fast, at least by a factor of $10^4$ with respect to chargino and
neutralinos, washing out the displaced vertex signal.

%%%%%%%%%%%%%%%%%%%%%%%%%%%Figure%%%%%%%%%%%%%%%%%%%%%%%%%%%%%%%%%
%%%%%%%%%%%%%%%%%%%%%%%%%%%Figure%%%%%%%%%%%%%%%%%%%%%%%%%%%%%%%%%

In order to be able to see the vertex, where the chargino or
neutralino decays to the secondary particles, we must require that the
decay products are such that this vertex can be reconstructed.  The
following decay modes allow us to reconstruct the neutralino decay
vertex
\begin{itemize}

  \item $\tilde{\chi}^0_1 \to \nu \ell^+ \ell^-$ with $\ell =e$, $\mu$
  denoted by $\ell \ell$;

  \item $\tilde{\chi}^0_1 \to \nu q \bar{q}$ denoted $jj$;

  \item $\tilde{\chi}^0_1 \to \tau q^\prime \bar{q}$, called $\tau jj$;

  \item $\tilde{\chi}^0_1 \to \nu b \bar{b}$, that we denote by $bb$;

  \item $\tilde{\chi}^0_1 \to \nu \tau^+ \tau^-$, called $\tau \tau$;

  \item  $\tilde{\chi}^0_1 \to \tau \nu  \ell$, called $\tau \ell$.

\end{itemize}
Clearly, the invisible decay of the neutralino into neutrinos cannot
be reconstructed and for this reason we discarded it.  On the other
hand, it is possible to measure the chargino decay vertex in the
decays:
\begin{itemize}

%\item $\tilde{\chi}^+_1 \to \nu \nu \ell^+ $ with $\ell =e$, $\mu$
%  denoted by $\nu\nu\ell$;

  \item $\tilde{\chi}^+_1 \to \nu q^{\prime} \bar{q}$ denoted $jj$;

  \item $\tilde{\chi}^+_1 \to \tau^+ q \bar{q}$, called $\tau jj$;

  \item $\tilde{\chi}^+_1 \to \tau^+ b \bar{b}$, called $\tau bb$;

  \item $\tilde{\chi}^+_1 \to \tau^+ \ell^+ \ell^-$, called $\tau\ell\ell$;

  \item $\tilde{\chi}^+_1 \to \ell^+ b \bar{b}$, that we denote by
    $\ell bb$;

\item $\tilde{\chi}^+_1 \to \ell^+ \ell^+ \ell^-$, that we denote by
  $\ell \ell\ell$;

\item $\tilde{\chi}^+_1 \to \ell^+ q \bar{q}$, that we denote by $\ell
  jj$. %;
 
%\item $\tilde{\chi}^+_1 \to \nu \nu \tau^+$ with the $\tau$ going into
%  a high multiplicity state. We call this mode by $\nu\nu\tau$.

\end{itemize}
Although the decay $\tilde{\chi}^+_1 \to \nu \nu \ell^+ $ can give
rise to a $e^\pm$ or $\mu^\pm$ with a high impact parameter, we do not
consider the mode in our study since it is not possible to obtain its
decay vertex.

We considered a crude model of the LHC detectors in order to identify
events containing displaced vertices.  We selected events presenting
neutralino or chargino decays away from the primary vertex via the
requirement that the displaced neutralino/chargino vertex is outside
an ellipsoid around the primary vertex
\[
      \left ( \frac{x}{\delta_{xy}} \right )^2
   +  \left ( \frac{y}{\delta_{xy}} \right )^2
   +  \left ( \frac{z}{\delta_{z}} \right )^2   = 1 \; ,
\]
where the $z$-axis is along the beam direction.  We took
$\delta_{xy}=100$ $\mu$m and $\delta_z = 2.5$ mm that correspond to
five times the expected resolution in each direction.  To guarantee a
high efficiency in the reconstruction of the displaced vertices
without a full detector simulation, we required the tracks leaving the
displaced vertex to be inside the rapidity coverage of the vertex
detector, {\em i.e.} $|\eta| < 2.5$.  Moreover, we also required that
the displaced vertices are inside the vertex detector -- that is, the
vertices must be within a radius of $550$ mm and z--axis length of
$800$ mm.

The SM backgrounds coming, for instance, from displaced vertices
associated to $b$'s or $\tau$'s can be eliminated by requiring that
the tracks defining a displaced vertex should have an invariant mass
larger than 20~GeV.  This way the displaced vertex signal passing all
the above cuts is essentially physics background free, however, there
might exist instrumental backgrounds which were not considered here.

An important issue in the displaced vertex search is the trigger on
the events containing them.  In order to mimic the triggers used by
the LHC collaborations, we accept events passing at least one of the
following requirements:
\begin{itemize}

\item The event has one isolated electron with $p_T > 20$ GeV;

\item The event has one isolated muon with $p_T > 6$ GeV;

\item The event has two isolated electrons with $p_T> 15$ GeV;

\item The event has one jet with $p_T > 100$ GeV;

\item The event has missing transversal energy $> 100$ GeV.

\end{itemize}

%%%%%%%%%%%%%%%%%%%%Section%%%%%%%%%%%%%%%%%%%%%%%%%%%%%%%%%
\section{Discovery Reach at the LHC}
\label{results}

%Here we present our results for the LHC reach in the search of BRpV
%mAMSB using the strategies outlined in Sec.~\ref{signals}.  
% In general our results exhibit little dependence on $m_0$ since both
% $m_{\chi_1^0}$ and $m_{\chi_1^+}$ are mainly determined by the
% gravitino mass with a mild dependence on $m_0$.
%
%OE:the above statement is not correct since the scalar masses change
% with m0. The same happens to production cross sections..

We estimated the LHC discovery reach in all the channels described in
Sec.~\ref{toplhc}. We required that either the signal leads to
$5\sigma$ departure from the background where this is not vanishing or
5 events in regions where there is no SM background.  We present our
results in the $m_{3/2} \otimes m_0$ plane for $\tan\beta=10$, $\mu>0$
and integrated luminosities of 10 fb$^{-1}$ and 100 fb$^{-1}$.

%%%

\begin{figure}[th]
  \begin{center}
  \includegraphics[width=7.75cm]{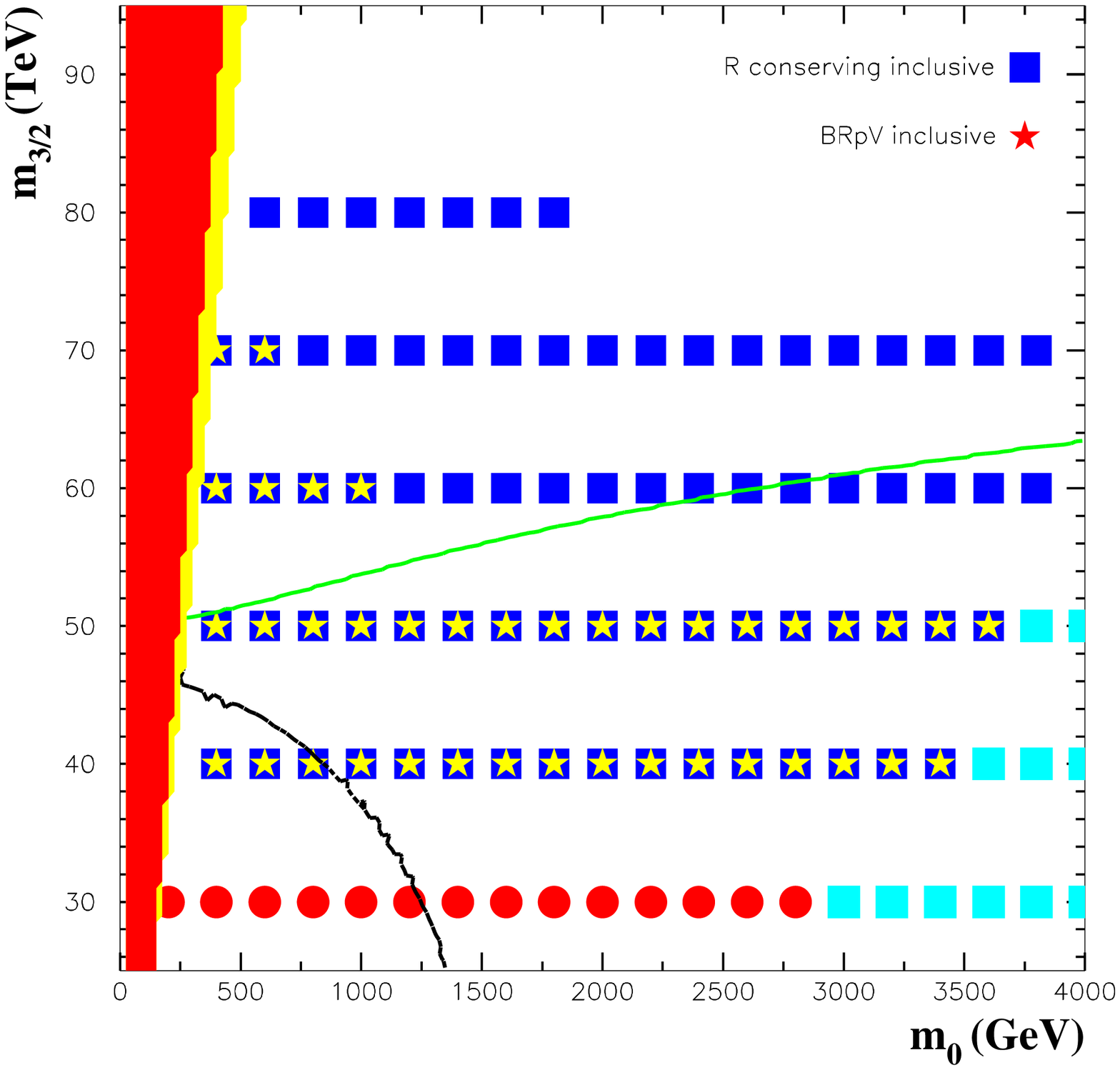}
  \includegraphics[width=7.75cm]{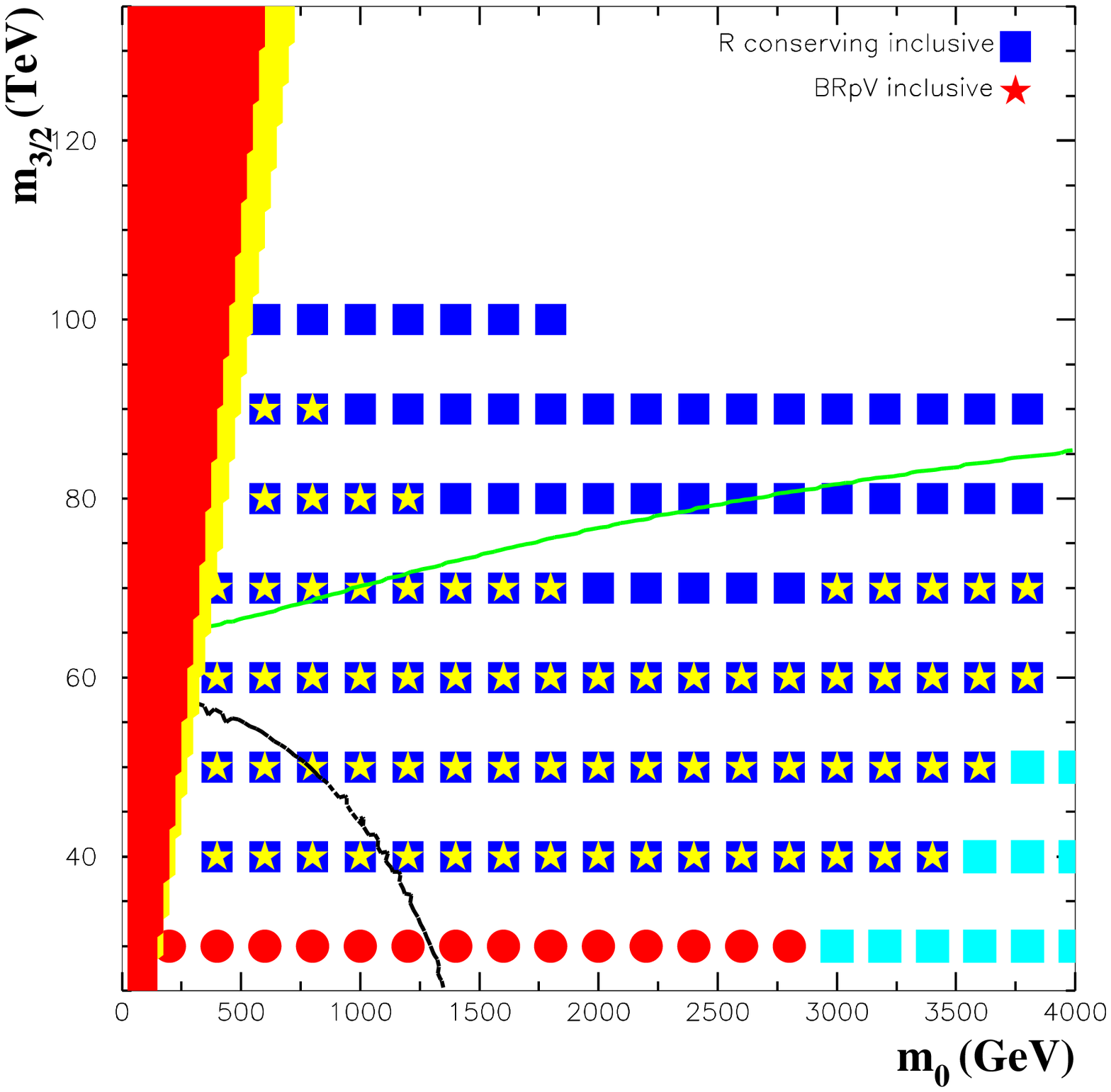}
  \end{center}
  \vspace*{-8mm}
  \caption{Discovery reach in the inclusive channel ({\bf IN}) for
    $\tan\beta=10$, $\mu>0$ and the integrated luminosities of 10
    fb$^{-1}$ (left panel) and 100 fb$^{-1}$ (right panel) . The dark
    (blue) square mark the points where mAMSB with R--parity
    conservation can be discovered while the stars stand for the reach
    of our BRpV--mAMSB model. Dark (red) circles stand for the points
    excluded by LEP data while no solution for the neutrino masses was
    found in the grey (cyan) squares.  The dark (red) area on the left
    is theoretically excluded by the existence of tachyonic particles
    and the light shaded area adjacent to this one is the area with a
    stau LSP.  The light (green) line is the contour for the gluino
    mass of 1 TeV. The Higgs mass is 114 GeV on the dashed dark line.
  }
\label{fig:di:in}
\end{figure}

%%%

We depict in Figure \ref{fig:di:in} the LHC discovery potential in the
all inclusive channel ({\bf IN}). For the sake of comparison, we also
present in this figure the discovery reach assuming R--parity
conservation. As we can see from this figure, the introduction of
bilinear R--parity violation reduces the reach in $m_{3/2}$ for a
given value of $m_0$ if we use the search strategy design for the
R--parity conserving case. Basically, the decay of the LSP reduces the
missing transverse energy making it harder to disentangle the SUSY
signal and the SM background. Comparing the left and right panels of
Fig.~\ref{fig:di:in}, we can see that a larger luminosity in this channel
expands the reach in $m_{3/2}$ from $\simeq 70$--80 TeV to 
$\simeq 90$--100 TeV.

%%%

\begin{figure}[th]
  \begin{center}
  \includegraphics[width=7.75cm]{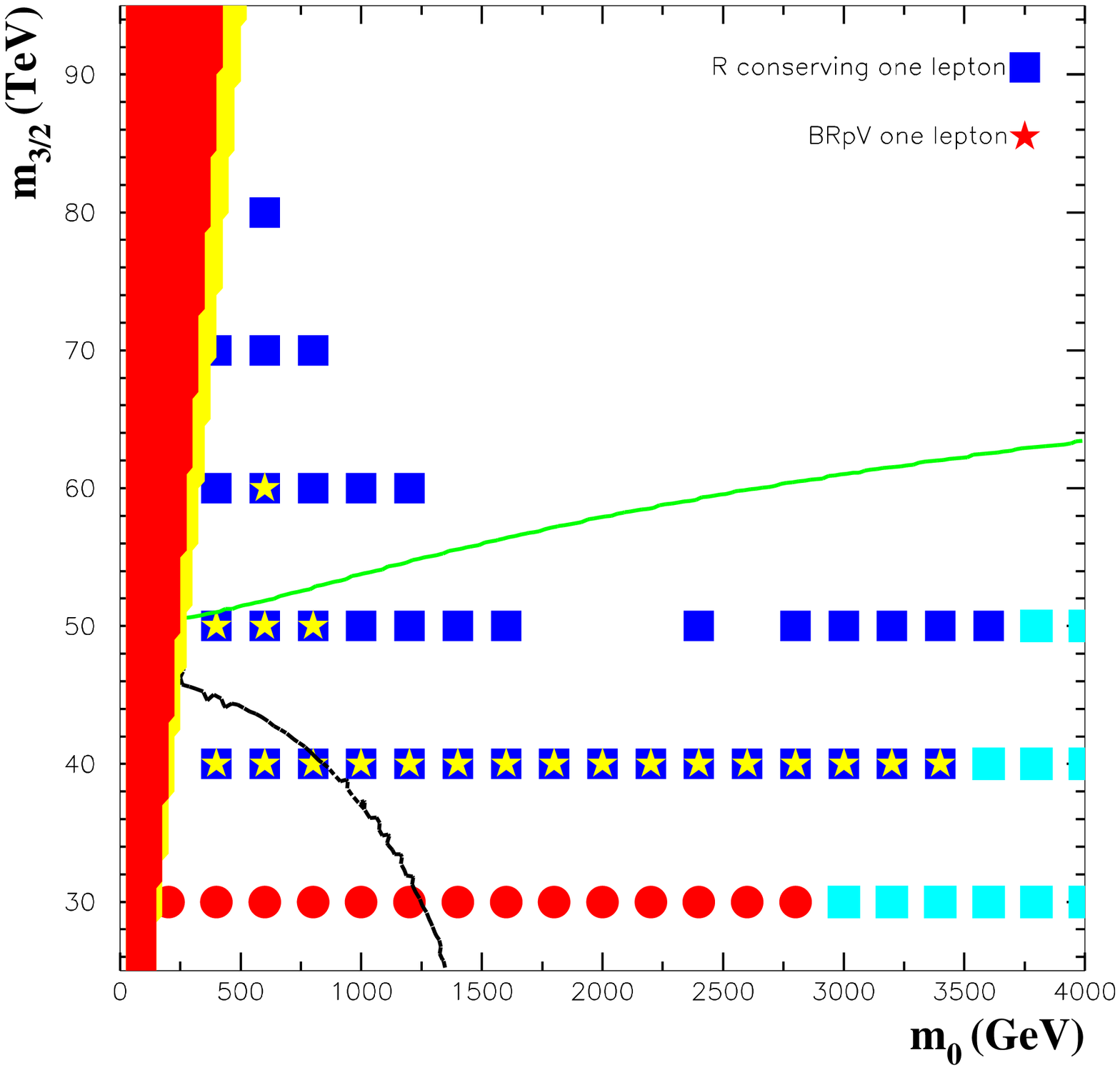}
  \includegraphics[width=7.75cm]{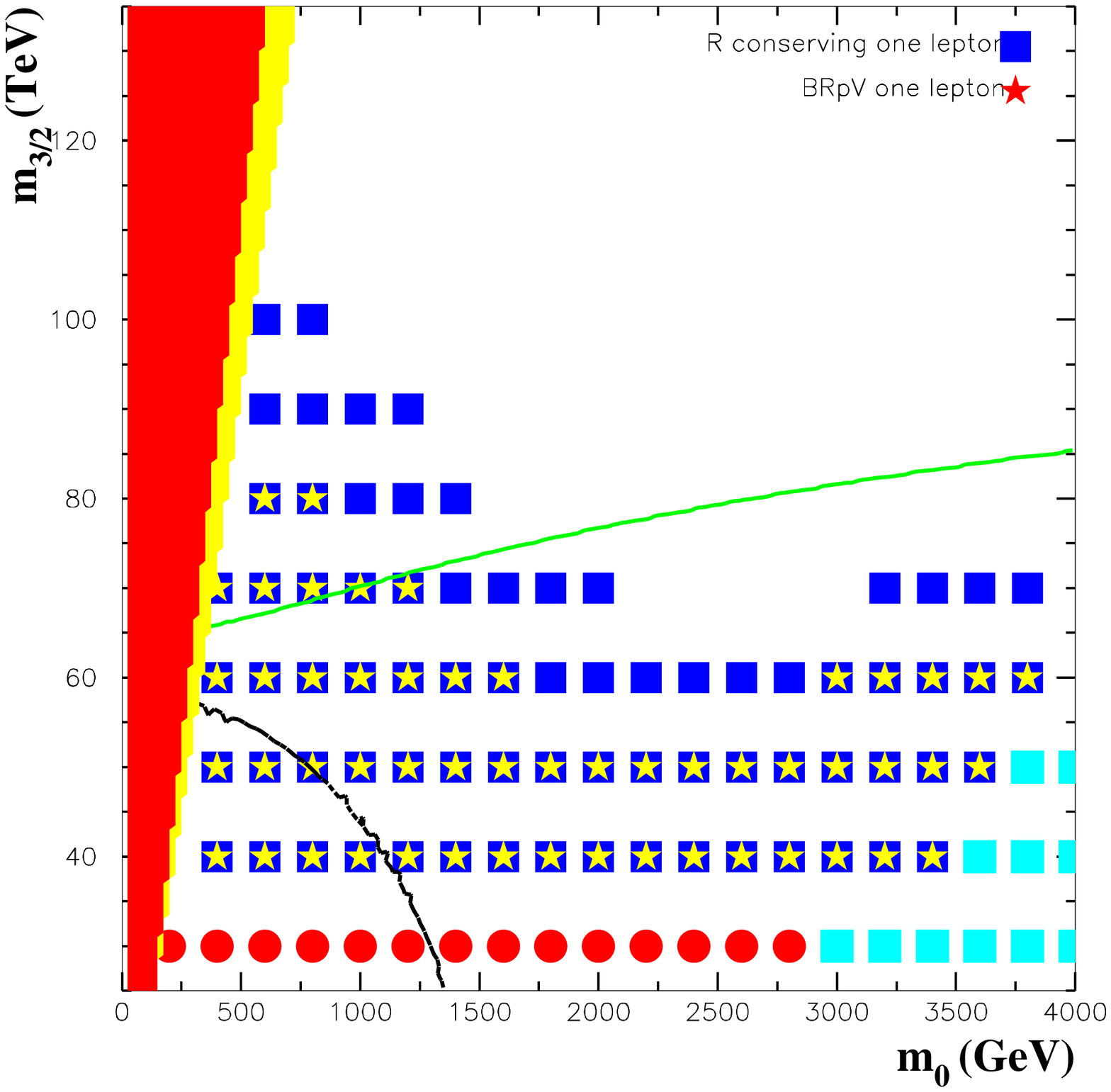}
  \end{center}
  \vspace*{-8mm}
  \caption{Discovery reach in the one lepton channel for the
    parameters and conventions used in Fig.~\ref{fig:di:in}.}
\label{fig:di:l1r}
\end{figure}

%%%

The reach in the {\bf 1}$\ell$ channel is presented in
Figure~\ref{fig:di:l1r}. In the R--parity conserving scenario this
channel is sensitive to the spectrum details of mAMSB since the
lightest chargino decay contains soft charged particles in addition to
the LSP and consequently, there are less channels available that give
rise to hard leptons.  Once again the introduction of R--parity
violation depletes this signal because of the reduced missing
transverse momentum. Moreover, the extra produced leptons in the
chargino or neutralino decays have the tendency of contributing to the
trilepton topology.  Comparing the left ad right panels of this figure
it is clear that the reach in this channel is extended with the
increase of the luminosity, however, its reach is still smaller than
the {\bf IN} one.

%%%

\begin{figure}[th]
  \begin{center}
  \includegraphics[width=7.75cm]{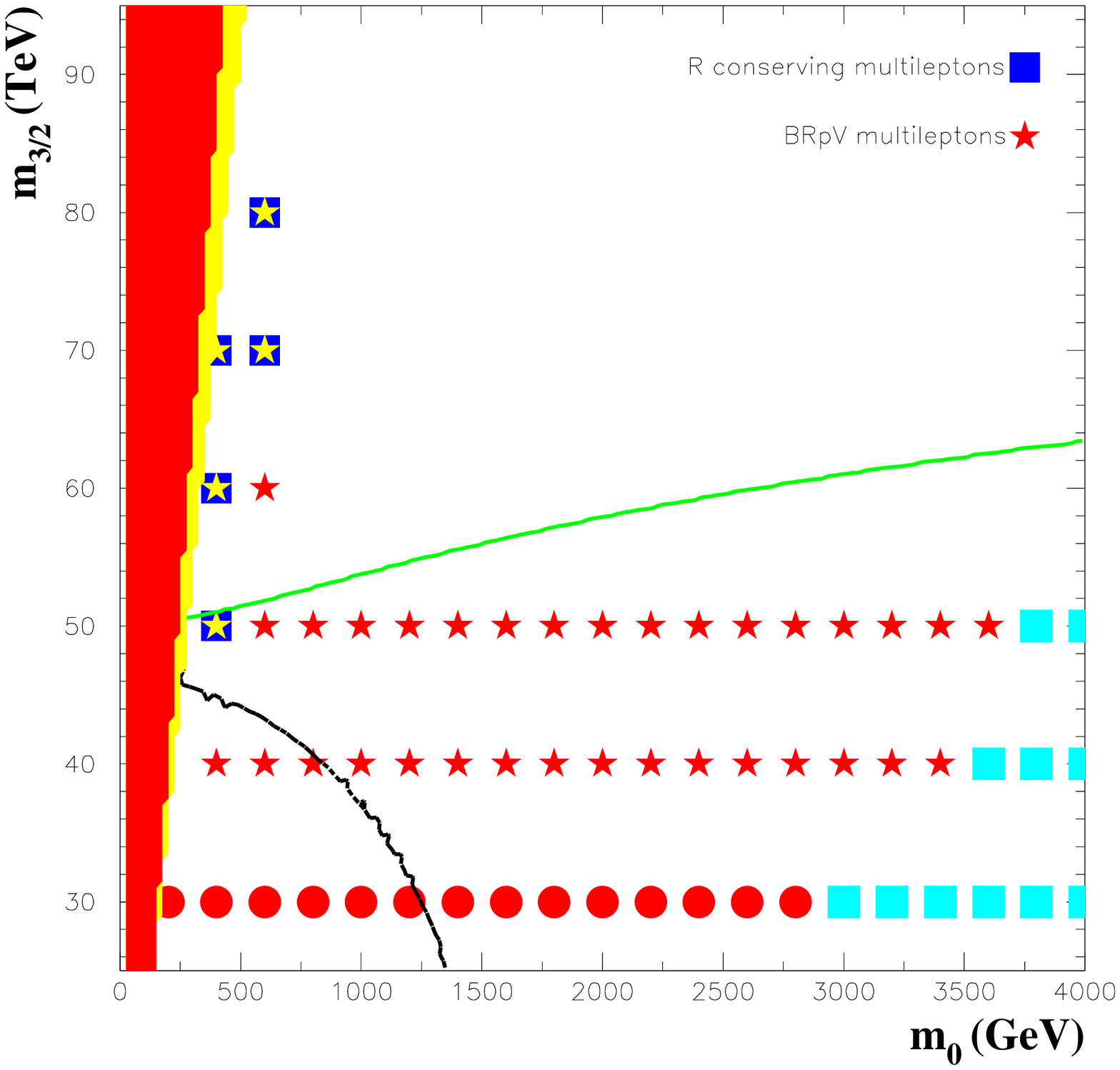}
  \includegraphics[width=7.75cm]{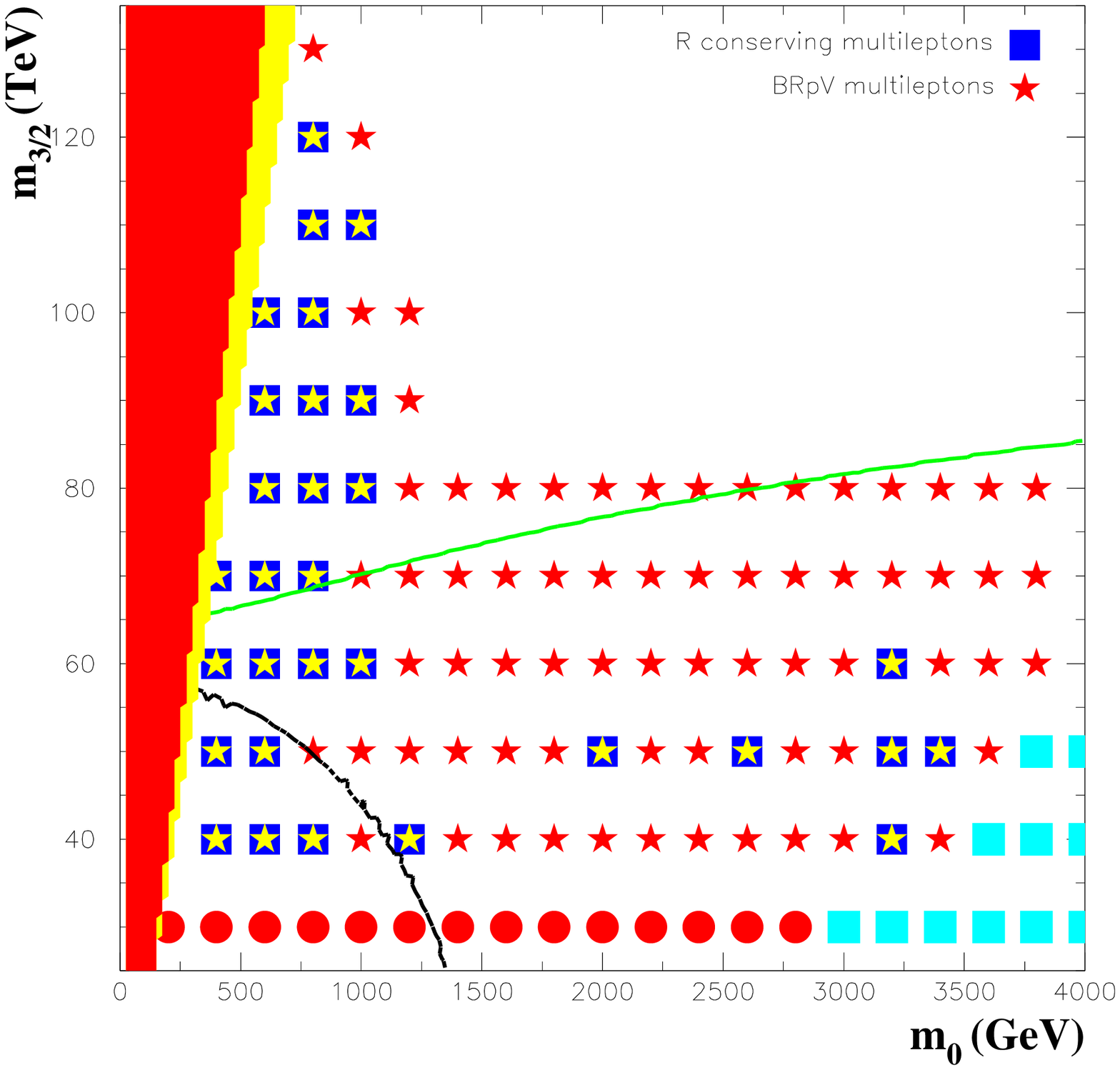}
  \end{center}
  \vspace*{-8mm}
  \caption{Discovery reach in the multilepton channel for the
    parameters and conventions used in Fig.~\ref{fig:di:in}.}
\label{fig:di:l3}
\end{figure}

%%%

The reach in the multilepton channel is depicted in
Figure~\ref{fig:di:l3}. For 10 fb$^{-1}$ the R--parity conserving
scenario has a very limited reach with the signal being sizeable only
in the area presenting light sleptons. The inclusion of BRpV
interactions increases the LHC reach in the $\mathbf{M\ell}$ channel
at small values of $m_{3/2}$ with the {\bf IN} and $\mathbf{M\ell}$
having similar reaches at low luminosities. At higher luminosities,
100 fb$^{-1}$, the multilepton channel reach is considerably extended
with and without R--parity conservation, being this the channel with
largest reach at small $m_0$. Moreover, this is the SUSY canonical
topology with the largest potential for discovery in the R--parity
violating scenario.

\begin{figure}[th]
   \begin{center} 
 \includegraphics[width=7.75cm]{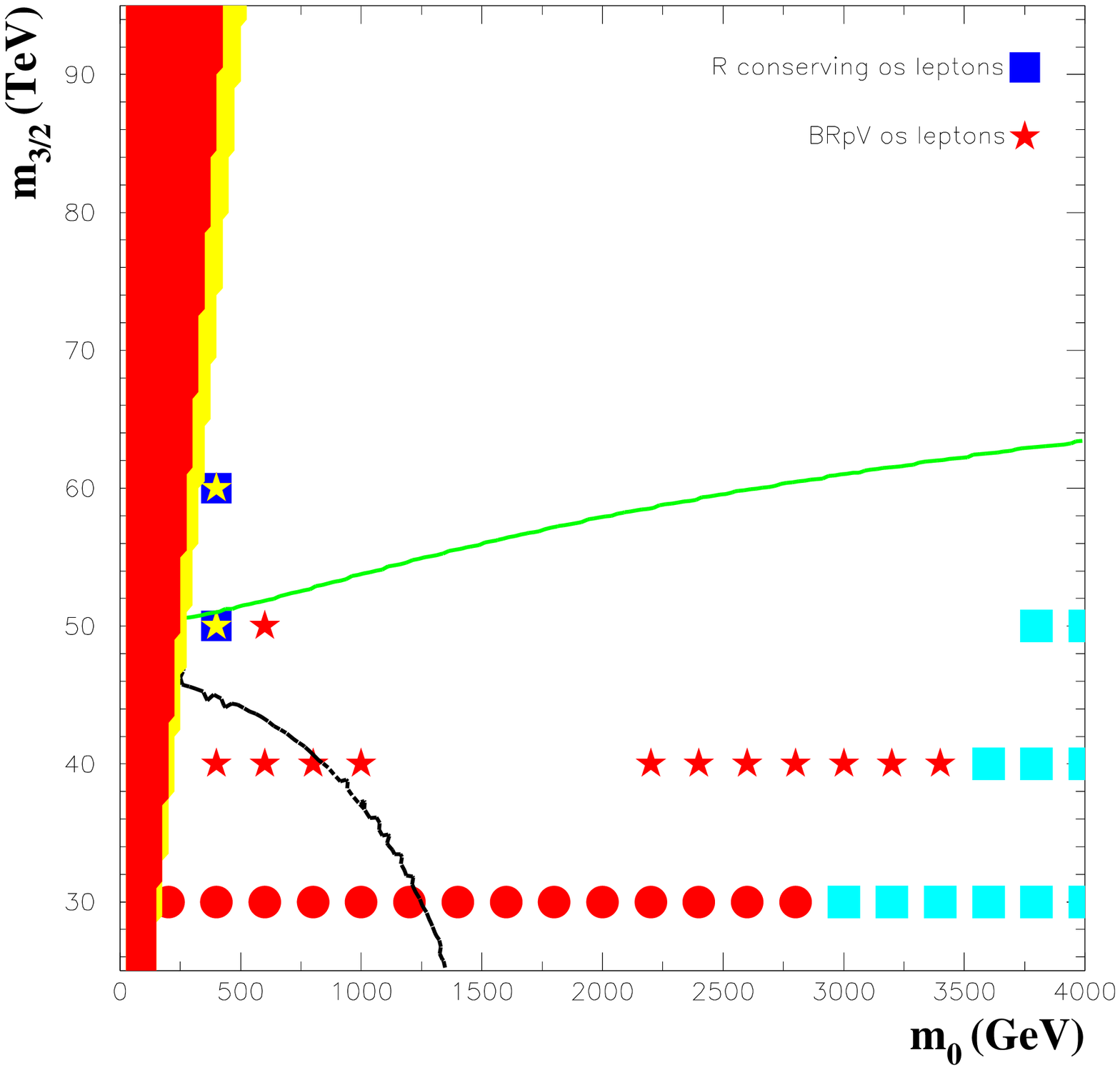} 
 \includegraphics[width=7.75cm]{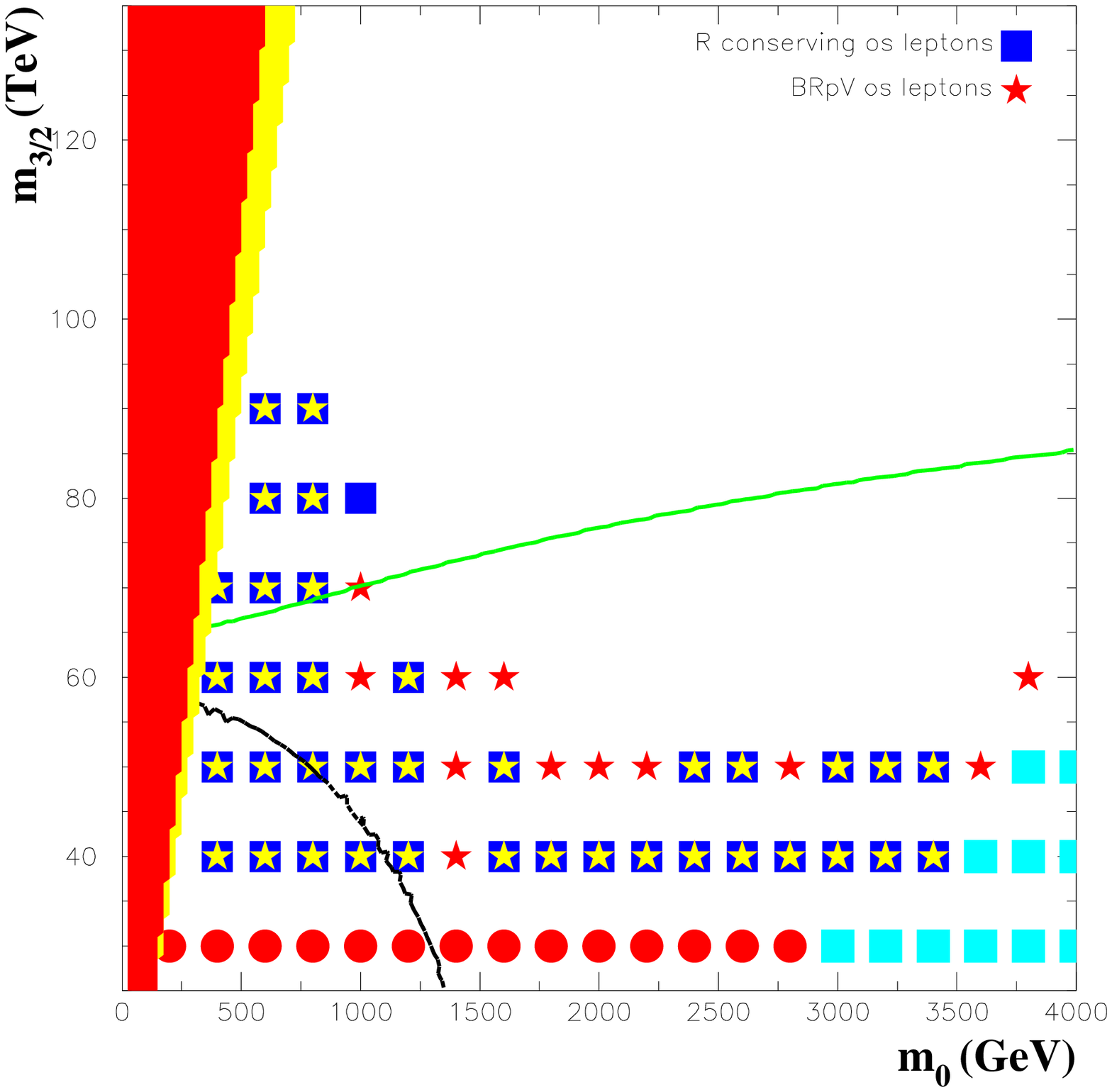} 
 \end{center}
   \vspace*{-8mm} \caption{Discovery reach in the opposite sign lepton channel
   for the parameters and conventions used in Fig.~\ref{fig:di:in}.}
 \label{fig:di:os}
 \end{figure}

\begin{figure}[th]
  \begin{center} 
\includegraphics[width=7.75cm]{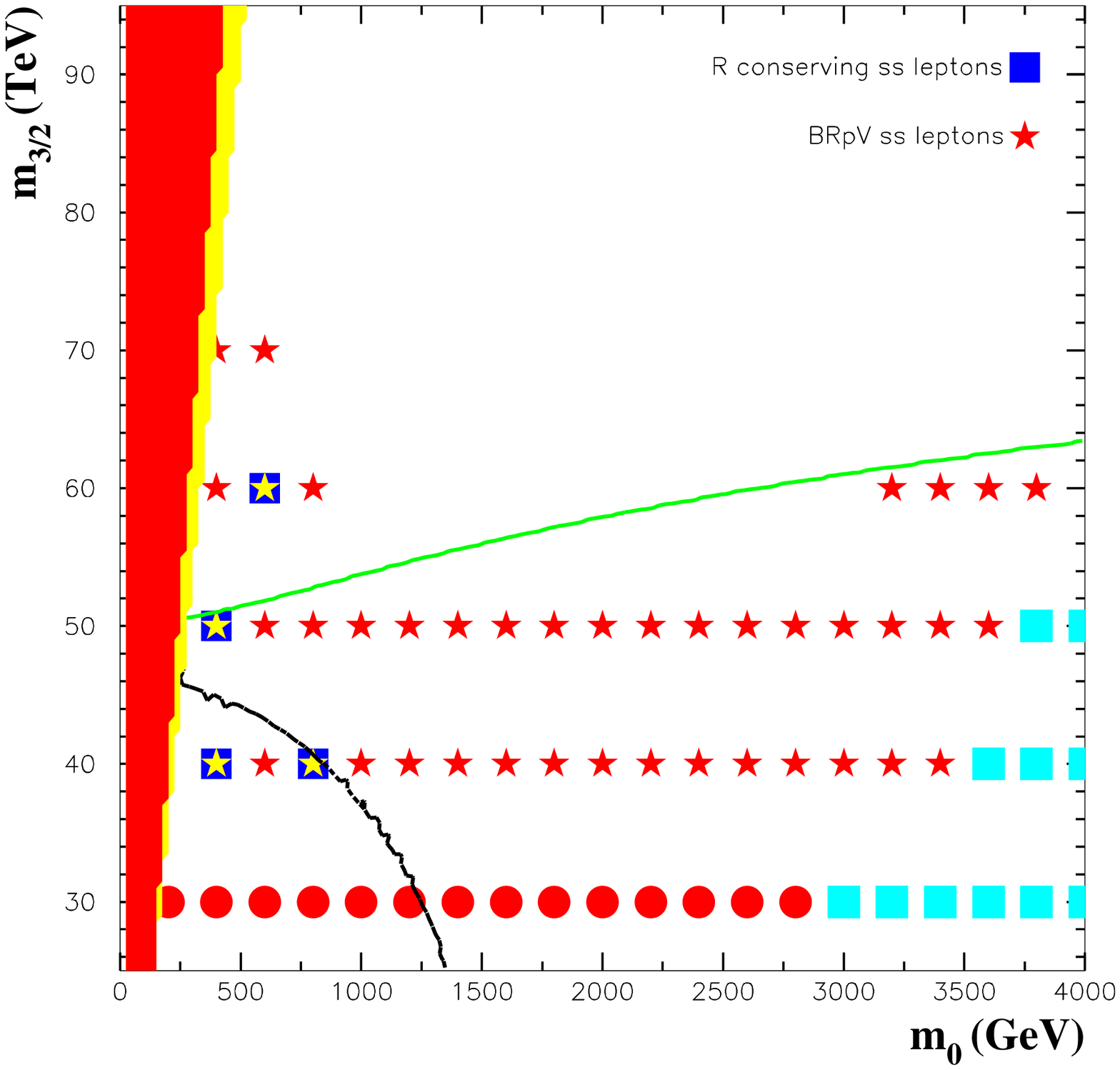} 
\includegraphics[width=7.75cm]{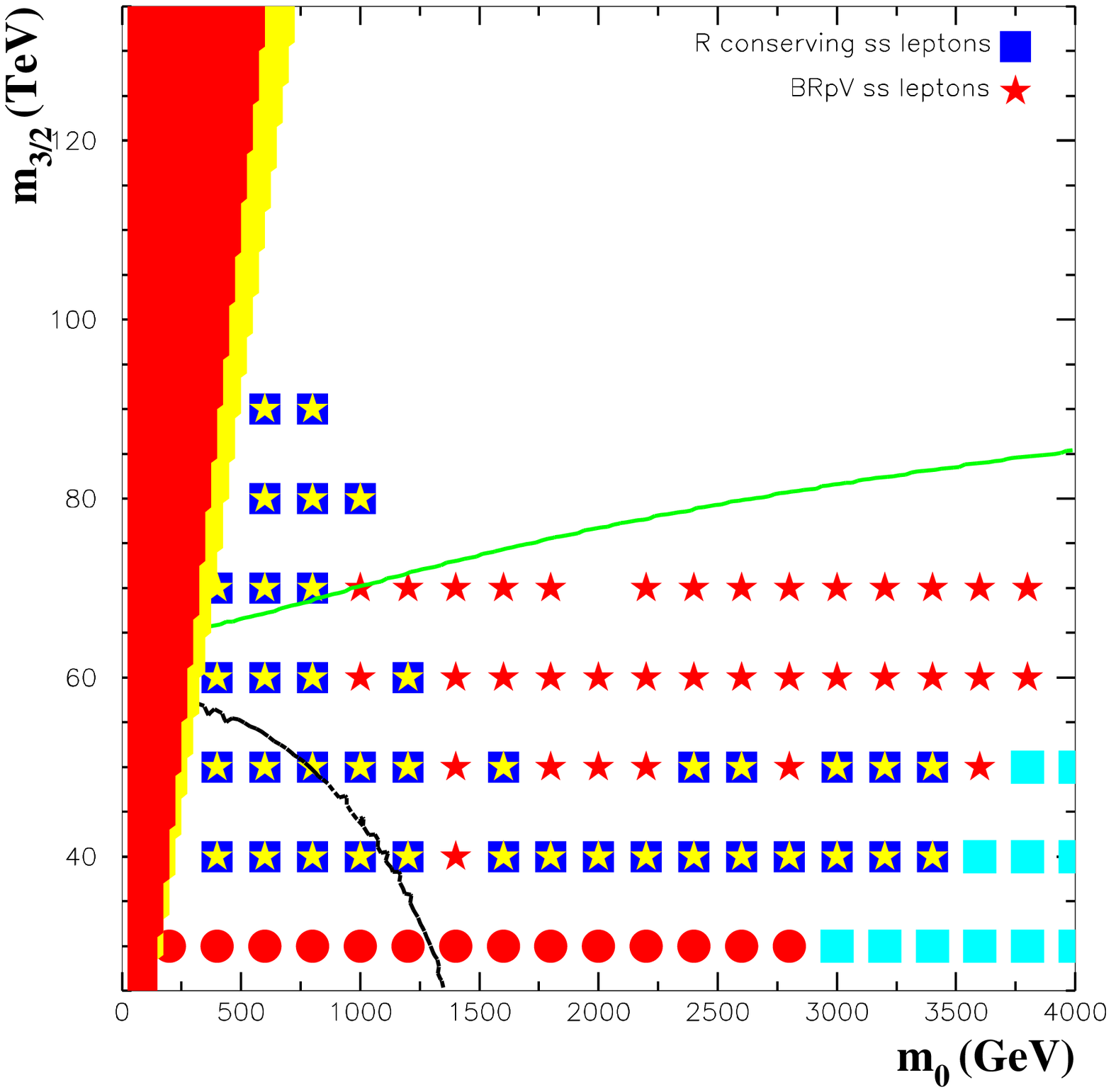} 
\end{center}
  \vspace*{-8mm} \caption{Discovery reach in the same sign lepton channel
  for the parameters and conventions used in Fig.~\ref{fig:di:in}.}
\label{fig:di:ss}
\end{figure}

We also studied the exclusive channels containing two isolated charged
leptons verifying that the introduction of R--parity violating
interactions enhances these channels.  In the R--parity violating
scenario the LHC {\bf OS} reach is similar to the $\mathbf{1\ell}$
channel one except at large $m_0$ where the {\bf OS} signal has a
reduced reach; see Figure~\ref{fig:di:os}. On the other hand, the {\bf
  SS} channel presents a smaller SM background in addition to an
enhanced signal due to the presence of majorana states in SUSY models.
Therefore, this topology has a good discovery reach in our BRpV-mAMSB
model as can be seen from Figure~\ref{fig:di:ss}. In fact, the {\bf SS}
final state has a slightly larger reach than the fully inclusive mode
{\bf IN}, with the {\bf SS} channel being the second most important
channel in the presence of BRpV interactions.

%%%
 
%%%

 Before we move on to the displaced vertex signal, we would like to
 stress that our results are an indication of R--parity violating
 interactions in the canonical SUSY searches. Certainly, the
 introduction of a larger number of floating cuts leads to larger
 reaches in all channels, like the ones in Ref.~\cite{Barr:2002ex}.
 Nevertheless, the depletion of the fully inclusive channel and the
 enhancement of exclusive topologies containing isolated leptons, as
 compared to the R--conserving case, must persist in a more elaborate
 analysis. In this light, it should be possible to determine whether R
 parity is broken or not by combining the results of all the canonical
 SUSY search channels.

%%%

%%%%%%%%%%%%%%%%%%%%%%%%%%%%%%%%%%%%%%%%%%%%%%%%%%%%%%%%%%%%%%%%%

\begin{figure}[th]
  \begin{center}
  \includegraphics[width=10.cm]{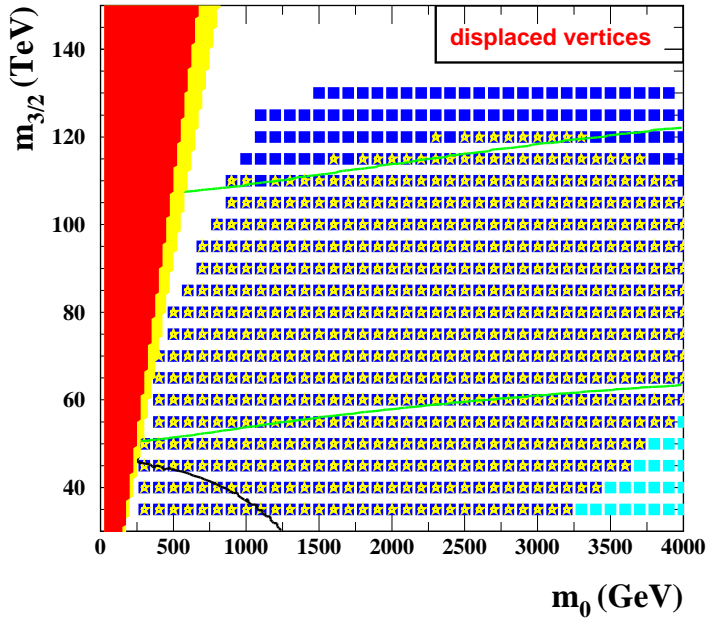}
  \end{center}
  \vspace*{-8mm}
  \caption{Discovery reach for the LHC in displaced vertices channel
    for $\tan\beta = 10$ and $\mu>0$ in the $m_{3/2}\otimes m_0$ plane 
    and an integrated luminosities of 10 fb$^{-1}$ (stars) and 
    100 fb$^{-1}$ (dark (blue) squares). The light (green) lines gives the 
    contour of a gluino mass of 1 TeV and 2 TeV. All other conventions 
    follow the ones of Fig.\ref{fig:di:in}.
     }
\label{dislhc}
\end{figure}

The smoking gun of BRpV-mAMSB is the existence of detached vertices
exhibiting a high invariant mass. There is no SM background for these
events except for possible instrumental backgrounds, rendering this
channel a very strong evidence for Physics beyond the standard model.
We present in Figure \ref{dislhc} the reach in the displaced vertex
topology for integrated luminosities of 10 and 100 fb$^{-1}$.  As we
can see from this figure, this channel does have the largest reach
($\simeq$ 110 (120) TeV for 10 (100) fb$^{-1}$) with the nice feature
of being almost independent of $m_0$, except at small $m_0$ where the
presence of light scalars lead to rapid neutralino and chargino decay;
see Fig.~\ref{fig:dl_10p}.

%%%%%%%%%%%%%%%%%%%%%%%%%%%%%%%%%%%%%%%%%%%%%%%%%%%%%%%%%%%%%%%%%%%%%%%%
%%%%%%%%%%%%%%%%%%%%%%%%%%%%%%%%%%%%%%%%%%%%%%%%%%%%%%%%%%%%%%%%%%%%%%

%%%%%%%%%%%%%%%%%%%%%%FIGURE%%%%%%%%%%%%%%%%%%%%%%%%%%%%%%%%%%%
\begin{figure}[thb]
  \begin{center}
  \includegraphics[width=10.1cm]{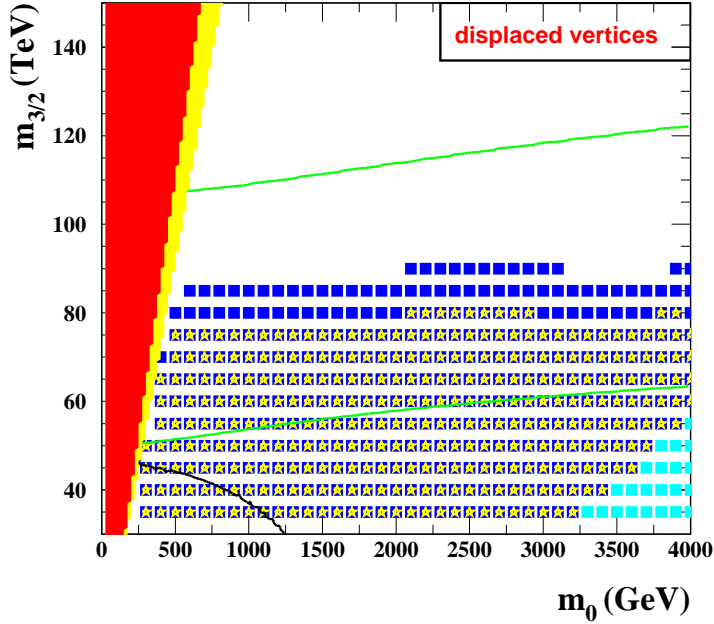}
  \end{center}
  \vspace*{-10mm}
  \caption{Discovery reach at the Tevatron in displaced vertices 
    channel for $\tan\beta = 10$, $\mu > 0$ in the
    $m_{3/2} \otimes m_0$ plane and an integrated luminosity 
    of 2 fb$^{-1}$ (stars) and 10 fb$^{-1}$ (dark (blue) squares). 
    The light (green) lines gives the contour of a gluino mass of 
    1 TeV and 2 TeV. All other conventions follow the ones of 
    Fig.\ref{fig:di:in}.
    }
\label{dis_tev}
\end{figure}

\section{Conclusions}

We have studied the phenomenology of AMSB model augmented with bilinear
R-parity parameters at the LHC. 
We show that the presence of bilinear R--parity interactions modifies
the canonical channels used for the supersymmetry search.  The decay
of the neutralino and chargino weaken the fully inclusive signal,
however, the existence of further leptons in the final state leads to
an enhancement of the leptonic exclusive modes.  In the case that the
BRpV-mAMSB final state contains three or more charged leptons the
reach is so enhanced that this channel alone has a comparable reach to
the fully inclusive one with R--parity conservation.  One interesting
aspect of the drastic change in the reach of the different topologies
is that a possible positive signal at the LHC can be used to
disentangle models with and without R--parity conservation.

Our BRpV-mAMSB model connects the R--parity violating parameters to
measured neutrino properties. Due to the smallness of the couplings
needed to reproduce the observed neutrino masses and mixings, our
model predicts a rather large lifetime for charginos and neutralinos.
This leads to the smoking gun of the BRpV-mAMSB theory, that is, the
existence of displaced vertices associated to neutralino or chargino
decays that exhibit a large visible invariant mass. We showed that the
search for such detached vertices does lead to the largest discovery
reach for such models. Indeed this channel can probe $m_{3/2}$ up to $
\simeq 100$ TeV, which constitutes almost all the natural parameter
space. At this point it is interesting to point out that the search of
detached vertices at the Tevatron can cover a large fraction of the
presently allowed parameter space. We depict in Figure~\ref{dis_tev}
the parameter space region that can be probed at the Tevatron for an
integrated luminosity of 2 fb$^{-1}$, showing that the Tevatron
collaborations can already cover a lot of ground in searching for 
BRpV-mAMSB.

%%%%%%%%%%%%%%%%%%%%%%%%%%%%%%%%%%%%%%%%%%%%%%%%%%%%%%%%%%%%
%\section{Conclusions}\label{conclusion}

\begin{acknowledgments}

  OE thanks the Kavli Institute for Theoretical Physics in Santa
  Barbara, CA for hospitality during the completion of this work. 
  SS thanks Funda\c{c}\~ao de Amparo \`a Pesquisa do Estado de S\~ao Paulo 
  (FAPESP) for financial support during the first part of this work.   
  We thank FAPESP and Conselho Nacional de Ci\^encia e Tecnologia (CNPq) 
  for financial support. This work was also supported by the German
  Ministry of Education and Research (BMBF) under contract 05HT6WWA,
  by the Chilean agency Conicyt grant No.~PBCT-ACT028, and by US
  department of energy under contract no.\ DE--FG03--91ER40674.

\end{acknowledgments}
%%%%%%%%%%%%%%%%%%%%%%%%%%%%%%%%%%%%%%%%%%%%%%%%%%%%%%%%%%%%
%\input{appA}
%%%%%%%%%%%%%%%%%%%%%%%%%%%%%%%%%%%%%%%%%%%%%%%%%%%%%%%%%%%%%%%%%%%%%%

%%%%%%%%%%%%%%%%%%%%%%%%%%%%%%%%%%%%%%%%%%%%%%%%%%%%%%%%%%%%%%%%%%%%%%
\end{document}